\documentclass[preprint,review,3p]{elsarticle}

\usepackage{algorithmic}
\usepackage{algorithm}

\usepackage{tabularx}
    \newcolumntype{L}{>{\raggedright\arraybackslash}X}
    
\usepackage{tikz}
\usepackage{pgfplots}
\pgfplotsset{compat=1.16}

\usepackage{caption}
\usepackage{subcaption}
\usepackage{moresize}
\usepackage{multirow}
\usepackage{url}

\usepackage[utf8]{inputenc}

\usepackage{xcolor}

\usepackage{graphicx}
\usepackage{amssymb}
\usepackage{amsthm,amsfonts,amsbsy,amsmath}
\DeclareMathAlphabet\mathbfcal{OMS}{cmsy}{b}{n}

\usepackage{lineno}

\journal{Signal Processing: Image Communication}

\begin{document}

\begin{frontmatter}


\title{Feature Fusion Via Dual-Resolution Compressive Measurement Matrix Analysis For Spectral Image Classification}



\author[mysecondaryaddress]{Juan Marcos Ramirez\corref{mycorrespondingauthor}}
\cortext[mysecondaryaddress]{Corresponding author}
\ead{juanmarcos.ramirez@urjc.es}

\author[mysecondaryaddress]{Jos\'e Ignacio Mart\'inez Torre}

\author[mymainaddress]{Henry Arguello}


\address[mysecondaryaddress]{Computer Science Department, Universidad Rey Juan Carlos, M\'ostoles, Spain}
\address[mymainaddress]{Computer Science Department, Universidad Industrial de Santander, Bucaramanga, Colombia}

\begin{abstract}
In the compressive spectral imaging (CSI) framework, different architectures have been proposed to recover high-resolution spectral images from compressive measurements. Since CSI architectures compactly capture the relevant information of the spectral image, various methods that extract classification features from compressive samples have been recently proposed. However, these techniques require a feature extraction procedure that reorders measurements using the information embedded in the coded aperture patterns. In this paper, a method that fuses features directly from dual-resolution compressive measurements is proposed for spectral image classification. More precisely, the fusion method is formulated as an inverse problem that estimates high-spatial-resolution and low-dimensional feature bands from compressive measurements. To this end, the decimation matrices that describe the compressive measurements as degraded versions of the fused features are mathematically modeled using the information embedded in the coded aperture patterns. Furthermore, we include both a sparsity-promoting and a total-variation (TV) regularization terms to the fusion problem in order to consider the correlations between neighbor pixels, and therefore, improve the accuracy of pixel-based classifiers. To solve the fusion problem, we describe an algorithm based on the accelerated variant of the alternating direction method of multipliers (accelerated-ADMM). Additionally, a classification approach that includes the developed fusion method and a multilayer neural network is introduced. Finally, the proposed approach is evaluated on three remote sensing spectral images and a set of compressive measurements captured in the laboratory. Extensive simulations show that the proposed classification approach outperforms other approaches under various performance metrics.
\end{abstract}

\begin{keyword}
Compressive spectral imaging \sep Dual-resolution acquisition systems \sep Feature fusion \sep Spectral image classification 



\end{keyword}

\date{August 22, 2020}

\end{frontmatter}


\section{Introduction}
\label{sec:intro}

Hyperspectral (HS) images are three-dimensional (3-D) datasets that capture the spectral information of bidimensional (2-D) scenes across a large number of spectral bands. These images have been used in different applications such as precision agriculture, urban planning, and disaster management  \cite{CampsRemote2011,BioucasHyperspectral2013}. In particular, the rich spectral information provided by HS images has been extensively considered to identify materials. However, HS sensors acquire low-spatial-resolution images to ensure measurements with a high signal-to-noise ratio (SNR) \cite{YokoyaHyperspectral2017}. On the other hand, multispectral (MS) images are high-spatial-resolution datasets with poor spectral information. Therefore, to exploit the information embedded in multi-resolution data, image fusion has emerged as a class of techniques that combines the information in HS and MS images for building high-resolution datasets \cite{YokoyaCoupled2012}.


Notice that the large sizes of both HS and MS images challenge the storing and processing capabilities of the acquisition systems. To overcome this drawback, compressive spectral imaging (CSI) has emerged as an alternative acquisition framework that captures the relevant information of spectral images by sampling a reduced number of measurements \cite{CaoComputational2016}. In this regard, the coded aperture snapshot spectral imaging (CASSI) is the most representative architecture whose functioning consists of capturing projections of an encoded and spectrally-dispersed version of the input image onto an imaging detector \cite{WagadarikarSingle2008}. Multiple CASSI based architectures have been proposed such as the three-dimensional CASSI (3D-CASSI) \cite{CaoComputational2016}, the colored CASSI \cite{ArguelloColored2014}, and the snapshot colored compressive spectral imager (SCCSI) \cite{correa2016multiple}. Moreover, various architectures that combine different kinds of measurements have been recently proposed to recover high-resolution spectral images \cite{ArguelloDual2015, JerezSingle2018,hauser2020dual}.  

Feature fusion from spectral images is a research field that focuses on exploiting the spectral and spatial information embedded in spectral images to improve identification and detection tasks. More precisely, this research area attempts to complement the spectral information with spatial features such as local smoothness, shape, and texture with the aim of improving the scene analysis \cite{imani2020overview}. In this context, feature fusion from CSI compressive measurements has become a challenging task due to the random nature of the coded aperture patterns and the nonlinearity of the encoding operation. In this sense, various feature fusion methods from CSI compressive measurements have been proposed. Specifically, a method that applies a compressive sensing reconstruction algorithm to recover the image PCA bands from CASSI measurements was developed in \cite{RamirezSpectral2014}. Recently, a feature extraction algorithm for spectral image classification from single-pixel measurements has been presented in \cite{VargasLow2019} using a low-rank matrix approximation model. Furthermore, various feature fusion methods from dual-resolution compressive measurements have been recently proposed \cite{HinojosaSpectral2019,RamirezMultiresolution2019,RamirezSpectral2020}. These methods exploit the fact that CSI systems can obtain the relevant information of the high-resolution image in a low dimensional space. Additionally, these approaches require a feature extraction stage that reorders measurements using the information embedded in the coded aperture patterns. More precisely, the method reported in \cite{HinojosaSpectral2019} stacks an interpolated version of the HS features and a superpixel map yielded by the MS features. Furthermore, the methods proposed in \cite{RamirezMultiresolution2019} and \cite{RamirezSpectral2020} fuse the extracted features by solving regularized optimization problems. Notice that the main difference with respect to the previous works relies on that the proposed approach obtains the fused features directly from compressive measurements without resorting to a feature extraction stage.

\begin{figure}
    \begin{center}
       \includegraphics[width=0.60\textwidth]{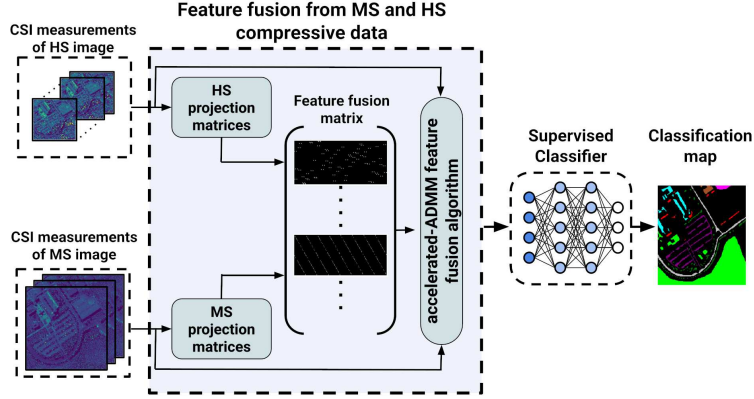}
    \end{center}
    \vspace{-15pt}
    \caption{Flowchart of the spectral image classification approach based on the proposed feature fusion method.}
    \vspace{-5pt}
    \label{fig:flowchart}
\end{figure}

This paper proposes a method that fuses features directly from HS and MS compressive measurements. In particular, the compressive measurements are obtained from a dual-arm optical architecture consisting of two 3D-CASSI branches. The contributions of this work are described as follows. 

\begin{enumerate}
\item First, we focus on deriving the expressions of the projection matrices that describe the compressive measurements as degraded versions of the fused features. Notice that these expressions incorporate the information embedded in the coded aperture patterns, therefore, a feature extraction procedure that reorders the measurements is not required.

\item From the observational model, the feature fusion is formulated as a regularized least-squares problem. Specifically, a sparsity-inducing term and a total-variation (TV) term are included in the cost function as regularizers. The sparsity-inducing term considers the spatial structures in spectral images while the TV term minimizes noise effects on the classification performance. Furthermore, an algorithm under an accelerated approach of the alternating direction method of multipliers \cite{OuyangAccelerated2015} is described to numerically solve the feature fusion problem.

\item  Finally, a deep multilayer neural network is used as a supervised pixel-based classifier for labeling high-resolution spectral images. 
\end{enumerate}


Figure \ref{fig:flowchart} illustrates the flowchart of the proposed feature fusion method in a spectral image classification framework. The proposed approach is evaluated on three datasets and a set of compressive measurements captured in the laboratory. Extensive simulations show that the proposed approach outperforms other methods that obtain features from dual-resolution compressive measurements. Furthermore, the developed method exhibits a remarkable performance when the projections are contaminated with additive noise. 

The paper is organized as follows. Section \ref{sec:dual} illustrates the acquisition system and the proposed feature fusion approach is developed in Section \ref{sec:methods2}. To evaluate the performance of the proposed feature fusion approach, the results of extensive simulations are shown in Section \ref{sec:results}. Finally, the concluding remarks are synthesized in Section \ref{sec:conclu}.

\section{Dual-arm optical system}
\label{sec:dual}

\begin{figure}
    \begin{center}
       \includegraphics[width=0.80\textwidth]{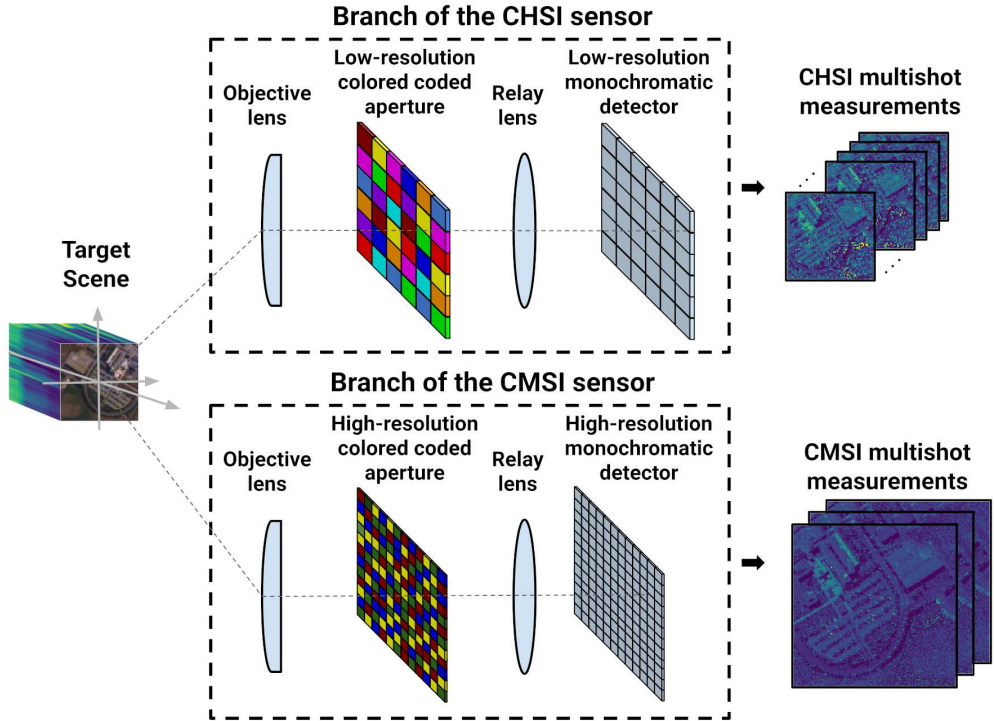}
    \end{center}
    \vspace{-10pt}
    \caption{Schematic of the dual-arm optical architecture.}
    \vspace{-5pt}
    \label{fig:dual_arm}
\end{figure}

In this work, a method that fuses features directly from dual-resolution compressive measurements is proposed for spectral image classification. More precisely, the proposed feature fusion method is developed for a dual-arm optical architecture, where each arm conssists of a multi-frame 3D-CASSI sensor. Figure \ref{fig:dual_arm} illustrates a schematic of the dual-arm optical architecture that obtains the compressive measurements. As can be seen in this figure, one branch is a 3D-CASSI sensor equipped with imaging lenses, a low-resolution colored coded aperture, and a low-resolution imaging detector. This branch, referred to as compressive hyperspectral imaging (CHSI), projects rich spectral information of the scene onto low-spatial resolution detectors. The second branch, referred to as compressive multispectral imaging (CMSI), is a 3D-CASSI sensor that includes imaging lenses, a high-spatial-resolution colored coded aperture, and a high-resolution camera detector. In general, this system obtains high-spatial-resolution snapshots with poor spectral information. Notice that the dual-arm optical systems have been widely used to acquire the high-resolution information of spectral images and spectral videos \cite{CSMUVI2012, RuedaDual2014, lin2014dual}. More precisely, these systems exploit the high and the low resolution coding masks to acquire dual-resolution compressive measurements, and thus, to capture the relevant information of the image of interest.


\subsection{Multi-frame 3D-CASSI architecture}
\label{sec:methods}

\begin{figure}
    \begin{center}
       \includegraphics[width=0.60\textwidth]{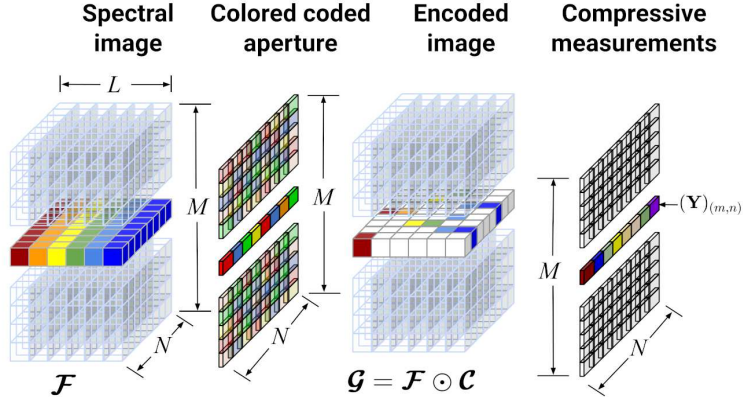}
    \end{center}
    \vspace{-10pt}
    \caption{Model of the acquisition process in the 3D-CASSI architecture.}
    \vspace{-5pt}
    \label{fig:3dcassi}
\end{figure}

Basically, the 3D-CASSI architecture projects an encoded version of the input image onto an imaging detector \cite{CaoComputational2016}. Figure \ref{fig:3dcassi} illustrates the model of the acquisition process. To describe 3D-CASSI measurements, consider $\mathbfcal{F}$ as a discrete version of the input image with a spatial resolution of $N \times M$ and $L$ spectral bands. Furthermore, let $\mathcal{F}_{(m,n,\ell)}$ be an element of the spectral image $\mathbfcal{F}$, where ($m$, $n$) denotes the spatial coordinate and $\ell$ represents the $\ell$-th spectral band. In this context, the input spectral image is encoded by a coded aperture that can be described as an array of optical filters with dimensions $M \times N$. Notice that each optical filter modulates the spectral signature at the spatial coordinate ($m$, $n$). In general, the coded aperture is modeled as a discrete cube $\mathbfcal{C}$ with dimensions $M \times N \times L$ and entries $\mathcal{C}_{(m,n,\ell)} \in \{0,1\}$. In consequence, the encoded image $\mathbfcal{G}$ is obtained by applying an element-wise product between the input image and the model of the coded aperture, i.e.  $\mathbfcal{G} = \mathbfcal{F} \odot \mathbfcal{C}$, where $\odot$ denotes the element-wise tensor multiplication. Subsequently, the encoded image is integrated along the sensor spectral sensitivity before it is projected onto a detector with dimensions $M \times N$ \cite{CaoComputational2016}. Assuming a perfect registration between each encoded spectral signature and the corresponding detector pixel, the intensity captured at the spatial coordinate ($m$, $n$) can be expressed as
\begin{equation}
    (\mathbf{Y})_{(m,n)} = \sum_{\ell=1}^{L} \mathcal{F}_{(m,n,\ell)} \mathcal{C}_{(m,n,\ell)},
\end{equation}
for $m = 1, \ldots, M$ amd $n = 1, \ldots, N$. 

Multiple snapshots are required to properly reconstruct the image from compressed measurements, where each snapshot is captured using a different coded aperture pattern. Various algorithms have been developed for designing coded aperture patterns under the restricted isometry property (RIP) constraint \cite{ArguelloColored2014, ParadaColored2017,HinojosaCoded2018}. In this work, we assume that a reduced set of optical filters are available with non-overlapping responses covering the entire wavelength spectrum. Furthermore, consider $\pmb{\Theta}$ a binary matrix with dimensions $L\times P$ whose columns $\{\pmb{\theta}_i \in \{0,1\}^L \}_{i=1}^{P}$ contain the responses of the available optical filters, where $P$ is the number of available optical filters. Therefore, the intensity captured by the imaging detector at the spatial location ($m$, $n$) and acquired at the $k$-th snapshot $(\mathbf{Y}^{(k)})_{(m,n)}$ can be described as
\begin{equation}
    (\mathbf{Y}^{(k)})_{(m,n)} = \sum_{\ell=1}^{L} \mathcal{F}_{(m,n,\ell)} \left(\pmb{\theta}_{\mathcal{S}_{(m,n,k)}}\right)_{\ell} 
\end{equation}
for $m = 1,\ldots,M$, $n=1,\ldots,N$, $k =1,\ldots,K$, where K denotes the number of captured snapshots, $\pmb{\theta}_{\mathcal{S}_{(m,n,k)}}$ represents the filter response used to encode the spectral signature at the coordinate ($m$, $n$) and the $k$-th snapshot with $\mathcal{S}_{(m,n,k)} \in \{1,2,\ldots,P\}$. To improve the performance of the spectral image recovery methods from multiple snapshots, the coded apertures should be designed such that projections capture the entire information of the spectral image \cite{ArguelloColored2014}. In this sense, Algorithm \ref{alg:ccad} has been introduced in \cite{RamirezMultiresolution2019} to optimally design coded apertures. Specifically, to capture the entire information of the image, each spectral signature should be modulated by all available filters. To this end, for every spatial coordinate ($m$,$n$), the design algorithm randomly distributes the optical filters across the snapshots. This distribution is included in $\boldsymbol{\alpha}$ that contains a random permutation of the set $\{1, \ldots, P \}$, where each component indexes a particular optical filter. In consequence, this algorithm generates a data cube $\mathbfcal{S}$ with dimensions $M \times N \times K$ that contains the filter patterns to capture measurements. The optical filter response used at the spatial location ($m$, $n$) and the snapshot $k$ is denoted as $\pmb{\theta}_{\mathcal{S}_{(m,n,k)}}$. In this work, we consider the information provided by the coded aperture patterns to build the decimation matrices of the proposed feature fusion algorithm.


\begin{algorithm}\small
\caption{Design of the colored coded apertures patterns \cite{RamirezMultiresolution2019}}\label{alg:ccad}
\begin{algorithmic}[1]
\renewcommand{\algorithmicrequire}{\textbf{Input:}}
\renewcommand{\algorithmicensure}{\textbf{Output:}}
\REQUIRE $\pmb{\Theta} \in \{0,1\}$
\ENSURE $\mathcal{S}_{(m,n,k)}$, $\pmb{\theta}_{\mathcal{S}_{(m,n,k)}}$ for $m = 1,\ldots,M$; $n=1,\ldots,N$; and $k=1,\ldots,K$ 
\FOR{$m = 1$ to $M$}
  \FOR{$n = 1$ to $N$}
    \STATE{$\boldsymbol{\alpha} =$ \texttt{randperm}($K$)}
      \FOR{$k = 1$ to $K$}
            \STATE{$\mathcal{S}_{(m,n,k)} = \boldsymbol{\alpha}(k)$}
           \STATE{$\pmb{\theta}_{\mathcal{S}_{(m,n,k)}} =\pmb{\theta}_{\boldsymbol{\alpha}(k)}$}
      \ENDFOR
  \ENDFOR
\ENDFOR
\end{algorithmic}
\end{algorithm}

\section{Feature fusion method}
\label{sec:methods2}

\subsection{High-resolution features}
\label{subsec:fused}


The fused features should exploit both the high-spatial-resolution acquired by the CMSI sensor and the rich spectral information captured by the CHSI system. In this work, each fused feature is modeled as the projections of the spectral signature of the high-resolution image modulated by the set of optical filters used by the CHSI sensor, in other words,
\begin{equation}\label{eq:hrf_model}
    \left(\mathbfcal{X}\right)_{(m,n,k)} = \sum_{\ell=1}^{L} \mathcal{F}_{(m,n,\ell)} \left(\pmb{\theta}^{(\mathtt{hs})}_{\mathcal{S}_{(m,n,k)}}\right)_{\ell} 
\end{equation}
for $k = 1,\ldots,K$, where $K$ is the number of captured snapshots, and $\{\pmb{\theta}^{(\mathtt{hs})}_{i}\}_{i=1}^{P^{(\mathtt{hs})}}$ are column vectors of the matrix $\pmb{\Theta}^{(\mathtt{hs})}$ that contains the optical filter responses used to obtain the HS compressive measurements. 


On the other hand, the CSI measurements do not offer discriminative properties required by pixel-based classification methods. This is because each CSI projection is acquired using a distinct coded aperture pattern. In previous works, a feature extraction procedure has been proposed before performing the feature fusion method \cite{HinojosaSpectral2019, RamirezMultiresolution2019, RamirezSpectral2020}. This feature extraction procedure reorders measurements such that each band corresponds to the spectral image response to a single optical filter. Conversely, in this work, we consider the information embedded in the coded aperture patterns to build the decimation matrices that describe the feature fusion model. Therefore, both feature extraction and feature fusion are included in a unique optimization problem.

\subsection{Multispectral downsampling model}
\label{subsec:multi_model}

\begin{figure}
\begin{center}
    \begin{tabular}{c c}
        \begin{subfigure}[b]{.45\textwidth}
            \begin{center}
            \includegraphics[width=\textwidth]{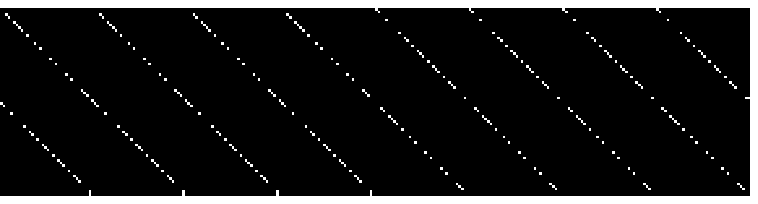}
            \end{center}
        \vspace{-15pt}
        \caption{}
        \label{subfig:ms_matrix}
        \end{subfigure}   
         &  
        \begin{subfigure}[b]{.45\textwidth}
            \begin{center}
            \includegraphics[width=\textwidth]{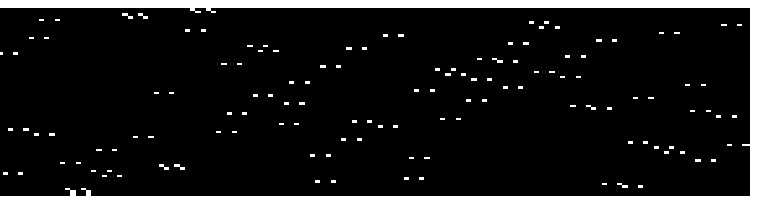}
            \end{center}
        \vspace{-15pt}
        \caption{}
        \label{subfig:hs_matrix}
        \end{subfigure}  
    \end{tabular}
\end{center}
\vspace{-20pt}
\caption{(a) An example of the projection matrix $\pmb{H}_{\mathtt{(ms)}}$ for $M=6$, $N=6$, $P=8$, and $q=4$; (b) an example of the projection matrix $\pmb{H}_{\mathtt{(hs)}}$ for $M=6$, $N=6$, $P=8$, and $p=2$.}
\vspace{-10pt}
\end{figure}

A CMSI measurement at the spatial coordinate ($m$, $n$) and the snapshot $w$ can be expressed as
\begin{equation}\label{eq:ms_model1}
    \left(\mathbfcal{X}_{\mathtt{(ms)}}\right)_{(m,n,w)} = \sum_{\ell=1}^{L} \mathcal{F}_{(m,n,\ell)} \left(\pmb{\theta}^{(\mathtt{ms})}_{\mathcal{S}_{(m,n,w)}}\right)_{\ell},
\end{equation}
for $w = 1,\ldots,W$, where $W$ represents the number of snapshots captured by the CMSI system, and  $\{\pmb{\theta}^{(\mathtt{ms})}_{i}\}_{i=1}^{P^{(\mathtt{ms})}}$ are columns of the matrix $\pmb{\Theta}^{(\mathtt{ms})}$ containing the responses of the optical filters used by the CMSI sensor. Without loss of generality, consider that the response of an optical filter used by a CMSI sensor, at the spatial coordinate ($m$,$n$) and the $w$-th snapshot, can be obtained as the summation of $q$ responses of optical filters used by a CHSI sensor, i.e.
\begin{equation}\label{eq:ms_degradation}
    \pmb{\theta}^{(\mathtt{ms})}_{\mathcal{S}_{(m,n,w)}} = \sum_{a=1}^{q} \pmb{\theta}^{(\mathtt{hs})}_{\mathcal{S}_{(m,n,(w-1)q+a)}}.
\end{equation}
Substituting (\ref{eq:ms_degradation}) in (\ref{eq:ms_model1}), we obtain
\begin{equation}\label{eq:ms_model2}
    \left(\mathbfcal{X}_{\mathtt{(ms)}}\right)_{(m,n,w)} = \sum_{a=1}^{q} \sum_{\ell=1}^{L} \mathcal{F}_{(m,n,\ell)} \left(\pmb{\theta}^{(\mathtt{hs})}_{\mathcal{S}_{\left(m,n,(w-1)q+a\right)}}\right)_{\ell}.
\end{equation}
As can be observed in (\ref{eq:ms_model2}), a CMSI measurement can be expressed as the linear expansion of $q$ components of the fused features, in other words, 
\begin{equation}\label{eq:ms_model3}
    \left(\mathbfcal{X}_{\mathtt{(ms)}}\right)_{(m,n,w)} = \sum_{a=1}^{q}  \left(\mathbfcal{X}\right)_{(m,n,(w-1)q+a)},
\end{equation}
for $m = 1,\ldots,M$; $n = 1,\ldots,N$; and $w = 1,\ldots,W$. Indeed, the model for the entire set of measurements can be synthesized in matrix form as follows
\begin{equation}\label{eq:ms_matrix}
\pmb{x}_{\mathtt{(ms)}} = \pmb{H}_{\mathtt{(ms)}} \pmb{x} + \boldsymbol{\eta}_{\mathtt{(ms)}}
\end{equation}
where $\pmb{x}_{\mathtt{(ms)}} \in \mathbb{R}^{MNW}$ are the CMSI measurements in vector form, $\pmb{H}_{\mathtt{(ms)}} \in \mathbb{R}^{MNW \times MNK}$ is the projection matrix that includes both downsampling operation and the information of the coded aperture patterns, $\pmb{x} \in \mathbb{R}^{MNK}$ is the fused feature vector, and $\boldsymbol{\eta}_{\mathtt{(ms)}} \in \mathbb{R}^{MNW}$ is the noise vector that contaminates the CMSI measurements. In general, the noise perturbations are characterized as random samples that follow a common statistical model. From the observational model (\ref{eq:ms_matrix}), each element of the projection matrix is obtained as
\begin{equation}\small
    (\pmb{H}_{\mathtt{(ms)}})_{(u,v)} = \left\lbrace 
    \begin{tabular}{c l}
        1, & if $u = m + (n-1)M + (w-1)MN$ and $v = m + (n-1)M + \left(\left( \mathcal{S}_{(m,n,w)}^{\mathtt{(ms)}} - 1\right)*q + z\right)MN$ \\ 
        0, & otherwise
    \end{tabular}
    \right.\label{eq:pmatrix_ms}
\end{equation}\normalsize
for $z=1,\ldots,q$, where $\mathbfcal{S}^{\mathtt{(ms)}}$ contains the coded aperture patterns generated to obtain the CMSI measurements. The projection matrix $\pmb{H}_{\mathtt{(ms)}}$ includes the information of the coded aperture patterns, therefore, the proposed method does not require a feature extraction procedure. Upon a closer examination of the sensing matrix dimensions, it can be observed that the measurement rate with respect to the fused features is reduced to $\varepsilon_{(\mathrm{ms})} = \frac{MN(K/q)}{MNK} =\frac{1}{q}$. Figure \ref{subfig:ms_matrix} illustrates an example of the projection matrix structure for $M=6$, $N=6$, $P=8$, and $q=4$.

\subsection{Hyperspectral downsampling model}
\label{subsec:hs_model}
On the other hand, a low-spatial-resolution measurement acquired by a CHSI system at the spatial coordinate ($\mu$,$\nu$) and snapshot $k$ can be described as
\begin{equation}\label{eq:hs_model1}
    \left(\mathbfcal{X}_{\mathtt{(hs)}}\right)_{(\mu,\nu,k)} = \sum_{\ell=1}^{L} \mathcal{F}_{(\mu,\nu,\ell)} \left(\pmb{\theta}^{\mathtt{(hs)}}_{\mathcal{S}_{(\mu,\nu,k)}}\right)_{\ell},
\end{equation}
with $\mu = 1,\ldots,M/p$ and $\nu = 1,\ldots,N/p$, where $p$ is the spatial decimation factor. In this case, each CHSI measurement can be obtained by averaging $p^2$ entries of the high-resolution features, this is,
\begin{equation}\label{eq:hs_model2}
    \left(\mathbfcal{X}_{\mathtt{(hs)}}\right)_{(\mu,\nu,k)} = \frac{1}{p^2} \sum_{a=0}^{p-1} \sum_{b=0}^{p-1}  \sum_{\ell=1}^{L} \mathcal{F}_{(m+a,n+b,\ell)} \left(\pmb{\theta}^{\mathtt{(hs)}}_{\mathcal{S}_{(\mu,\nu,k)}}\right)_{\ell},
\end{equation}
therefore, a CHSI sample can be expressed as
\begin{equation}\label{eq:hs_model3}
    \left(\mathbfcal{X}^{\mathtt{(hs)}}\right)_{(\mu,\nu,k)} = \frac{1}{p^2} \sum_{a=0}^{p-1} \sum_{b=0}^{p-1}  \left(\mathbfcal{X}\right)_{(m+a,n+b,k)}.
\end{equation}

The model for the entire set of CHSI measurements in matrix form can be compactly described as
\begin{equation}\label{eq:hs_matrix}
\pmb{x}_{\mathtt{(hs)}} = \pmb{H}_{\mathtt{(hs)}} \pmb{x} + \boldsymbol{\eta}_{\mathtt{(hs)}}
\end{equation}
where $\pmb{x}_{\mathtt{(hs)}} \in \mathbb{R}^{(M/p)(N/p)K}$ is the vector that contains the CHSI measurements, $\pmb{H}_{\mathtt{(hs)}} \in \mathbb{R}^{(M/p)(N/p)K \times MNK}$ is the projection matrix that includes the spatial downsampling, $\pmb{x} \in \mathbb{R}^{MNK}$ are the fused features in vector form, and $\boldsymbol{\eta}_{\mathtt{(hs)}} \in \mathbb{R}^{(M/p)(N/p)K}$ is the additive noise vector corrupting the CHSI measurements. Notice that the projection matrix $\pmb{H}_{\mathtt{(hs)}}$ also includes the information of the coded aperture patterns generated to capture the CHSI samples, therefore, a feature extraction stage is not required.  Particularly, each component ($u$,$v$) of the projection matrix can be obtained as
\begin{equation}\small
    (\pmb{H}_{\mathtt{(hs)}})_{(u,v)} = \left\lbrace 
    \begin{tabular}{c l}
        $\frac{1}{p^2}$, & if $u = \mu + (\nu-1)(M/p) + (k-1)(M/p)(N/p)$ and $v = \mu + (\nu-1)(M/p) + tM + z + d$ \\ 
        0, & otherwise
    \end{tabular}
    \right.\label{eq:pmatrix_hs}
\end{equation}\normalsize
for $z=1,\ldots,p$ and $z=1,\ldots,p$; where $d = \lfloor \frac{\mu p}{M} \rfloor + \mathrm{mod}(\mu, (M/p))p +  \left(\mathcal{S}_{(\mu,\nu,t)}^{\mathtt{(hs)}}-1\right)MN$, $\mathbfcal{S}^{\mathtt{(hs)}}$ contains the coded aperture patterns used for acquiring the CHSI measurements. Furthermore, $\lfloor\cdot\rfloor$ is the floor operator that extracts the integer part of its argument, $\mathrm{mod}(\mu,(M/p))$and outputs the reminder after division between $\mu$ and $M/p$. In this case, the measurement rate with respect to the fused features is reduced to $\varepsilon_{(\mathrm{hs})} = \frac{(M/p)(N/p)K}{MNK} = \frac{1}{p^2}$. Figure \ref{subfig:hs_matrix} displays an example of the projection matrix structure for $M=6$, $N=6$, $P=8$, and $p=2$.

\subsection{Feature fusion algorithm}
\label{subsec:problem}

Let assume the entries of $\pmb{\eta}_{(\mathtt{ms})}$ and $\pmb{\eta}_{(\mathtt{hs})}$ are described as random samples that follow a Gaussian distribution with variance $\sigma_{(\mathtt{ms})}^2$ and $\sigma_{(\mathtt{hs})}^2$, respectively, for $a=1,\ldots,MNW$ and $b=1,\ldots,(M/p)(N/p)K$. Furthermore, without loss of generality, we assume that $\sigma_{(\mathtt{ms})}^2 = \sigma_{(\mathtt{hs})}^2$. This leads to the minimization of the sum of squared errors as the optimization approach to describe the fusion problem. Two regularization terms are aggregated to this formulation to improve the quality of the fused features. The feature fusion problem is formulated as
\begin{equation}
    \hat{\pmb{x}} = \arg\min_{\pmb{x}} \left\lbrace \frac{1}{2} \|\pmb{y} - \pmb{H} \pmb{x}  \|_2^2 + \lambda_1 \|\boldsymbol{\Psi}^{\top} \pmb{x} \|_1 + \lambda_2\|\pmb{x}\|_{TV} \right\rbrace,
\end{equation}
where $\lambda_1$ and $\lambda_2$ are the regularization parameter that control the trade-off between the sum of squared errors and the regularization terms, with 
\begin{eqnarray}
\pmb{y} & = & [\pmb{x}_{ms}^{\top}, \pmb{x}_{hs}^{\top}]^{\top}, \\
\pmb{H} & = & [\pmb{H}_{ms}^{\top}, \pmb{H}_{hs}^{\top}]^{\top}.
\end{eqnarray}


As shown in Section \ref{subsec:fused}, each band of the high-resolution features can be considered as the response of the input spectral density to a particular optical filter. To preserve the spatial structure of the scene, the sparsity-inducing term $\lambda_1 \|\boldsymbol{\Psi}^{\top}\pmb{x}\|_1$ is included. In other words, we assume that the merged features can be sparsely described in an orthogonal transform $\boldsymbol{\Psi}$. This approach has been widely exploited in image denoising, improving thus, the quality of the recovered images. On the other hand, the total-variation (TV) term $\lambda_2\|\pmb{x}\|_{TV}$  exploits the local spatial information present in spectral image features. Specifically, the used TV norm is described as follow
\begin{equation}
    \| \pmb{x} \|_{TV}  = \sum_{(i,j,k)} \left\lbrace |x_{(i,j,k)} - x_{(i+1,j,k)}| + |x_{(i,j,k)} - x_{(i,j+1,k)}| + |x_{(i,j,k)} - x_{(i,j,k+1)}| \right\rbrace,
\end{equation}
where $x_{(i,j,k)}$ is the spectral image voxel at the location ($i,j,k$). The total variation norm is typically rewritten as
\begin{equation}
    \| \pmb{x} \|_{TV} = \| \boldsymbol{\Phi} \pmb{x} \|_1 
\end{equation}
where $\boldsymbol{\Phi}$ is the first-order difference operator. This term induces piece-wise spatial smoothness on the recovered features preserving, in turn, the edges of the scene \cite{IordacheTotal2012}. It is worth noting that TV norm has been considered in various feature extraction methods to improve the classification accuracy \cite{RastiHyperspectral2016,rasti2019hyperspectral}.

The feature fusion problem can be also described as
\begin{equation} \label{eq1}
\begin{split}
\hat{\pmb{x}} & = \arg\min_{\pmb{x}} \left\lbrace \frac{1}{2} \|\pmb{y} - \pmb{H} \pmb{x}  \|_2^2 + \lambda_1 \|\pmb{\gamma}_1 \|_1 + \lambda_2\|\pmb{\gamma}_{\textcolor{blue}{2}}\|_{1} \right\rbrace \\
 & \mathrm{s.t.}~\boldsymbol{\Psi}^{\top}\pmb{x} - \pmb{\gamma}_1 = 0,~\boldsymbol{\Phi}\pmb{x} - \pmb{\gamma}_2 = 0,
\end{split}
\end{equation}
with $\pmb{\gamma}_1$ and $\pmb{\gamma}_2$ as auxiliary variables in a alternating direction optimization approach. Thus, the augmented Lagrangian for (\ref{eq1}) can be expressed as
\begin{equation}\label{eq:augm}
    \mathbfcal{L}(\pmb{x},\pmb{\gamma}_1,\pmb{\gamma}_2) = \frac{1}{2} \|\pmb{y} - \pmb{H} \pmb{x}  \|_2^2 + \lambda_1 \|\pmb{\gamma}_1 \|_1 + \lambda_2\|\pmb{\gamma}_2\|_{1} + \frac{\rho}{2} \|\boldsymbol{\Psi}^{\top}\pmb{x} - \pmb{\gamma}_1 - \pmb{\delta}_1 \|_2^2 + \frac{\rho}{2} \|\boldsymbol{\Phi}\pmb{x} - \pmb{\gamma}_2 - \pmb{\delta}_2 \|_2^2,
\end{equation}
where $\rho > 0$ is a penalty parameter, and ($\pmb{\delta}_1$ , $\pmb{\delta}_2$) are Lagrange multiplier vectors. To estimate the fused features, the minimization of the augmented Lagrangian is performed. Algorithm \ref{alg:fusion} displays the pseudocode to fuse features using the accelerated approach of ADMM \cite{ouyang2015accelerated}. Notice that we aim at minimizing (\ref{eq:augm}) with respect to $\pmb{x}$, $\pmb{\gamma}_1$, and $\pmb{\gamma}_2$. As can be seen in steps 4, 7, and 8 of Algorithm \ref{alg:fusion}, the update of each variable is computed as a weighted sum of previous estimates ($\pmb{x}^{(\jmath)}$, $\pmb{\gamma}_1^{(\jmath)}$, $\pmb{\gamma}_2^{(\jmath)}$) and current approximations ($\pmb{x}^{(\jmath +1)}$, $\pmb{\gamma}_1^{(\jmath+1)}$, $\pmb{\gamma}_2^{(\jmath+1)}$). To compute the current estimate $\pmb{x}^{(\jmath +1)}$, we focus on minimizing the augmented Lagrangian with respect to the target variable. Under this approach, the estimation of $\pmb{x}$ at the iteration $(\jmath + 1)$ is obtained by solving the minimization problem
\begin{equation}
\pmb{x}^{(\jmath + 1)}  \longleftarrow \arg\min_{\pmb{x}} \left\lbrace
\frac{1}{2} \|\pmb{y} - \pmb{H} \pmb{x}  \|_2^2 + \frac{\rho}{2} \|\boldsymbol{\Psi}^{\top}\pmb{x} - \pmb{\gamma}_1^{(\jmath)} - \pmb{\delta}_1^{(\jmath)} \|_2^2 + \frac{\rho}{2} \|\boldsymbol{\Phi}\pmb{x} - \pmb{\gamma}_2^{(\jmath)} - \pmb{\delta}_2^{(\jmath)} \|_2^2 
\right\rbrace.  \label{eq:xmin_admm}
\end{equation}

Notice that the analytical solution of (\ref{eq:xmin_admm}) involve computationally expensive operations, therefore, a rough estimation is obtained by using the first-order approximation of the cost function around $\pmb{x}^{(\jmath)}$, in other words,

\vspace{-10pt}
\small
\begin{eqnarray}
\frac{1}{2} \|\pmb{y} - \pmb{H} \pmb{x}  \|_2^2 + \frac{\rho}{2} \|\boldsymbol{\Psi}^{\top}\pmb{x} - \pmb{\gamma}_1^{(\jmath)} - \pmb{\delta}_1^{(\jmath)} \|_2^2 + \frac{\rho}{2} \|\boldsymbol{\Phi}\pmb{x} - \pmb{\gamma}_2^{(\jmath)} - \pmb{\delta}_2^{(\jmath)} \|_2^2 & \approx &
\frac{1}{2} \|\pmb{y} - \pmb{H} \pmb{x}^{(\jmath)}  \|_2^2 + \frac{\rho}{2} \|\boldsymbol{\Psi}^{\top}\pmb{x}^{(\jmath)} - \pmb{\gamma}_1^{(\jmath)} - \pmb{\delta}_1^{(\jmath)} \|_2^2 \nonumber \\
& & + \frac{\rho}{2} \|\boldsymbol{\Phi}\pmb{x}^{(\jmath)} - \pmb{\gamma}_2^{(\jmath)} - \pmb{\delta}_2^{(\jmath)} \|_2^2 +\left\langle \pmb{x} - \pmb{x}^{(\jmath)}, \vartheta(\pmb{x}^{(\jmath)})  \right\rangle \nonumber \\
& & + \frac{\beta}{2}\|\pmb{x} - \pmb{x}^{(\jmath)}\|_2^2  \nonumber \\
& \approx & \frac{1}{2} \|\pmb{y} - \pmb{H} \pmb{x}^{(\jmath)}  \|_2^2 + \frac{\rho}{2} \|\boldsymbol{\Psi}^{\top}\pmb{x}^{(\jmath)} - \pmb{\gamma}_1^{(\jmath)} - \pmb{\delta}_1^{(\jmath)} \|_2^2 \nonumber \\
& & + \frac{\rho}{2} \|\boldsymbol{\Phi}\pmb{x}^{(\jmath)} - \pmb{\gamma}_2^{(\jmath)} - \pmb{\delta}_2^{(\jmath)} \|_2^2 \nonumber \\
& & + \frac{\beta}{2}\|\pmb{x} - \pmb{x}^{(\jmath)} + \frac{1}{\beta} \vartheta(\pmb{x}^{(\jmath)}) \|_2^2, \label{eq:apprx0}
\end{eqnarray}\normalsize 
where $\vartheta(\pmb{x}^{(\jmath)})$ is the gradient of the cost function around $\pmb{x}^{(\jmath)}$ that is computed as follows
\begin{equation}\label{eq:grad}
   \vartheta(\pmb{x}^{(\jmath)}) = \pmb{H}^{\top} \left(\pmb{y} - \pmb{H}\pmb{x}^{(\jmath)} \right) + \rho\left(\pmb{x}^{(\jmath)} - \boldsymbol{\Psi}^{\top} \left(\pmb{\gamma}_1^{(\jmath)} - \pmb{\delta}_1^{(\jmath)}\right)\right) + \rho\left(\boldsymbol{\Phi}^{\top}\left(\boldsymbol{\Phi}\pmb{x}^{(\jmath)} - \pmb{\gamma}_2^{(\jmath)} - \pmb{\delta}_2^{(\jmath)}\right)\right),
\end{equation}
with $\beta$ as a step parameter. By substituting (\ref{eq:grad}) in (\ref{eq:apprx0}), and setting the derivative to zero, the current approximation $\pmb{x}^{(\jmath+1)}$ is obtained by

\vspace{-10pt}
\small
\begin{equation}
\pmb{x}^{(\jmath+1)} = \pmb{x}^{(\jmath)} - \left(\frac{1}{\beta}\right)\left(\pmb{H}^{\top} \left(\pmb{y} - \pmb{H}\pmb{x}^{(\jmath)} \right) + \rho\left(\pmb{x}^{(\jmath)} - \boldsymbol{\Psi}^{\top} \left(\pmb{\gamma}_1^{(\jmath)} - \pmb{\delta}_1^{(\jmath)}\right)\right) + \rho\left(\boldsymbol{\Phi}^{\top}\left(\boldsymbol{\Phi}\pmb{x}^{(\jmath)} - \pmb{\gamma}_2^{(\jmath)} - \pmb{\delta}_2^{(\jmath)}\right)\right) \right). \label{eq:updatex}
\end{equation}\normalsize

Additionally, to update the auxiliary variables $\pmb{\gamma}_1$ and $\pmb{\gamma}_2$ at the iteration ($\jmath + 1$), the optimization problems should be solved
\begin{equation}
\pmb{\gamma}_1^{(\jmath + 1)}  \longleftarrow \arg\min_{\pmb{\gamma}_1} \left\lbrace
\frac{\rho}{2} \|\boldsymbol{\Psi}^{\top}\pmb{x}^{(\jmath +1)} - \pmb{\gamma}_1 - \pmb{\delta}_1^{(\jmath)} \|_2^2 + + \lambda_1 \|\pmb{\gamma}_1 \|_1 
\right\rbrace  
\end{equation}
\begin{equation}
\pmb{\gamma}_2^{(\jmath + 1)}  \longleftarrow \arg\min_{\pmb{\gamma}_2} \left\lbrace
\frac{\rho}{2} \|\boldsymbol{\Phi}\pmb{x}^{(\jmath + 1)} - \pmb{\gamma}_2 - \pmb{\delta}_2^{(\jmath)} \|_2^2 + \lambda_2\|\pmb{\gamma}_2\|_{1}
\right\rbrace  
\end{equation}
whose solutions are obtained basically applying soft-threholding operators, as shown as follows 
\begin{eqnarray}
    \pmb{\gamma}_1^{(\jmath+1)} & = & \mathrm{soft}\left(\boldsymbol{\Psi}^{\top} \pmb{x}^{(\jmath+1)} + \boldsymbol{\delta}_1^{(\jmath)},\lambda_1 / \rho \right) \label{eq:lmbd1} \\
    \pmb{\gamma}_2^{(\jmath+1)} & = & \mathrm{soft}\left(\boldsymbol{\Phi} \pmb{x}^{(\jmath+1)} + \boldsymbol{\delta}_2^{(\jmath)},\lambda_2 / \rho \right) \label{eq:lmbd2}
\end{eqnarray}
where $\mathrm{soft}(\gamma,\lambda) = \mathrm{sign}(\gamma)\max (\gamma - \lambda,0)$. Then, Lagrange multiplier vectors are updated by computing the operations described in steps 9 and 10 of Algorithm \ref{alg:fusion}. Finally, the computational complexity for computing $\pmb{x}$ can be determined as $\mathcal{O}((MN)^2 K)$, while the soft-thresholding operations exhibit a complexity $\mathcal{O}(MNK)$, with $M\times N \times K$ as the size of the fused features. In consequence, the computational complexity per iteration is obtained as $\mathcal{O}((MN)^2 K)$. The source code for the proposed feature fusion method can be downloaded from this link: \url{https://github.com/JuanMarcosRamirez/featurefusion_getfund}

\begin{algorithm}
\caption{Accelerated ADMM for feature fusion from compressive measurements}\label{alg:fusion}
\begin{algorithmic}[1]\small
\renewcommand{\algorithmicrequire}{\textbf{Input:}}
\renewcommand{\algorithmicensure}{\textbf{Output:}}
\REQUIRE $\pmb{y}$, $\boldsymbol{\Psi}$, $\boldsymbol{\Phi}$, $\pmb{H}$, $\lambda_1$, $\lambda_2$, $\rho$ $\alpha^{(0)}$ \newline
$\pmb{x}^{(0)} \rightarrow \mathbf{0}$, $\pmb{\gamma}_1^{(0)} \rightarrow \mathbf{0}$, $\pmb{\gamma}_2^{(0)} \rightarrow \mathbf{0}$, $\pmb{\delta}_1^{(0)} \rightarrow \mathbf{0}$, $\pmb{\delta}_2^{(0)} \rightarrow \mathbf{0}$
\ENSURE $\pmb{x}^{(\jmath+1)}$
\REPEAT
    \STATE{$\pmb{x}^{(\jmath)} =\left(1-\alpha^{(\jmath + 1)}\right)\pmb{x}_2^{(\jmath)} + \alpha^{(\jmath + 1)} \pmb{x}^{(\jmath)}$} 
    \STATE{$\pmb{x}^{(\jmath+1)} \leftarrow \arg\min_{\pmb{x}} \mathbfcal{L}(\pmb{x},\pmb{\gamma}_1^{(\jmath)},\pmb{\gamma}_2^{(\jmath)}) \mathtt{~~Eq. (\ref{eq:updatex})}$}
    \STATE{$\pmb{x}_2^{(\jmath + 1)} =\left(1-\alpha^{(\jmath + 1)}\right)\pmb{x}_2^{(\jmath)} + \alpha^{(\jmath + 1)} \pmb{x}^{(\jmath+1)}$} 
    \STATE{$\pmb{\gamma}_1^{(\jmath+1)} \leftarrow \arg\min_{\pmb{\gamma}_1} \mathbfcal{L}(\pmb{x}^{(\jmath)},\pmb{\gamma}_1,\pmb{\gamma}_2^{(\jmath)}) \mathtt{~~Eq. (\ref{eq:lmbd1})}$}
    \STATE{$\pmb{\gamma}_2^{(\jmath+1)} \leftarrow \arg\min_{\pmb{\gamma}_2} \mathbfcal{L}(\pmb{x}^{(\jmath)},\pmb{\gamma}_1^{(\jmath)},\pmb{\gamma}_2)\mathtt{~~Eq. (\ref{eq:lmbd2})}$}
    \STATE{$\pmb{\gamma}_1^{(\jmath + 1)} =\left(1-\alpha^{(\jmath + 1)}\right)\pmb{\gamma}_1^{(\jmath)} + \alpha^{(\jmath + 1)} \pmb{\gamma}_1^{(\jmath+1)}$}
    \STATE{$\pmb{\gamma}_2^{(\jmath + 1)} =\left(1-\alpha^{(\jmath + 1)}\right)\pmb{\gamma}_2^{(\jmath)} + \alpha^{(\jmath + 1)} \pmb{\gamma}_2^{(\jmath+1)}$} 
    \STATE{$\boldsymbol{\delta}_1^{(\jmath + 1)} = \boldsymbol{\delta}_1^{(\jmath)} + \boldsymbol{\Psi}^{\top} \pmb{x}^{(\jmath +1)} - \pmb{\gamma}_1^{(\jmath +1)}$}
    \STATE{$\boldsymbol{\delta}_2^{(\jmath +1)} = \boldsymbol{\delta}_2^{(\jmath)} + \boldsymbol{\Phi} \pmb{x}^{(\jmath +1)} - \pmb{\gamma}_2^{(\jmath +1)}$}
\UNTIL a stopping rule is satisfied
\end{algorithmic}
\end{algorithm}

\subsection{Supervised classification method}\label{subsec:classification}

The feature fusion method is included in a spectral image classification approach as shown in Fig. \ref{fig:flowchart}. To this end, a multilayer perceptron neural network (MLPNN) is selected as a supervised classification method. More precisely, an MLPNN is a feed-forward neural network whose structure can learn nonlinear classification boundaries \cite{goodfellow2016deep}. This classifier typically exhibits a deep structure that can handle large data sets and it does not require a huge amount of training data for achieving the desired classification performance compared to that needed by the convolutional neural networks (CNN). Notice that various methods based on deep learning have been recently developed for spectral image classification exhibiting remarkable results \cite{mei2019spectral, LiDeep2019}. However, these methods require a high-resolution spectral image. Furthermore, these techniques involve complex network structures that consider separately the spectral signature and local spatial information. Indeed, to achieve the desired classification performance, both a huge amount of training data and a large number of learning epochs are needed.

\begin{figure}
    \begin{center}
       \includegraphics[width=0.70\textwidth]{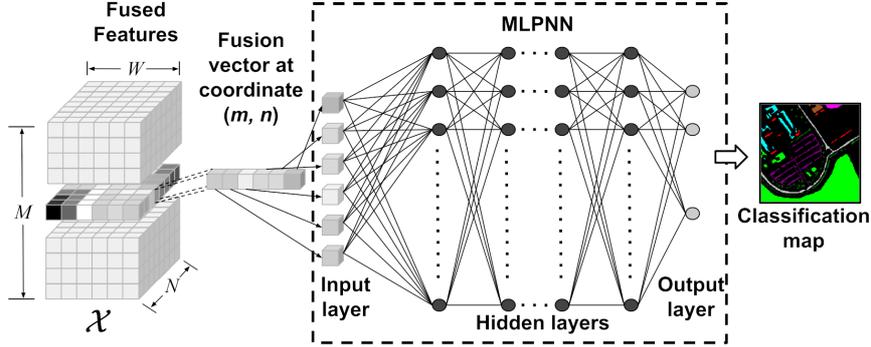}
    \end{center}
    \vspace{-10pt}
    \caption{Schematic of the supervised classification approach.}
    \vspace{-5pt}
    \label{fig:ffnn}
\end{figure}

A schematic of the supervised classification approach is illustrated in Fig. \ref{fig:ffnn}. As can be observed in this figure, the MLPNN consists of an input layer, a set of hidden layers, and an output layer. Specifically, the proposed classification approach firstly extracts the vector from the fused features at the coordinate ($m$, $n$) and its elements reduce to the entries of the MLPNN input layer. Notice that the size of the input layer depends on the dimensions of the fused vectors, and consequently, it also depends on the compression ratio. Then, the MLPNN comprises a set of fully-connected hidden layers whose $j$-th neuron at the $k$-th layer outputs
\begin{equation}
    \zeta_{j}^{k} = f\left(\sum_{i=1}^{N_n} \Upsilon_{i,j}^{k-1} \zeta_{i,j}^{k - 1} + \xi_{j}^{k - 1} \right)
\end{equation}
where $\Upsilon_{i,j}^{k-1}$ is the weight that connects the $i$-th neuron at the $k-1$ layer with the $j$-th neuron in layer $k$, $\xi_{j}^{k - 1}$ is the bias term of the $j$-th neuron in layer $k$, and $f(\cdot)$ represents a nonlinear activation function. In this work, we select ten hidden layers with ten neurons each, where the number of hidden layers and the number of neurons by layer were determined by implementing a classification performance search. Every neuron is activated with the rectified linear unit (ReLU) function, i.e. $f(u) = \mathrm{max}(u,0)$. The output layer size depends on the number of the classes and the network parameters are updated by using the backpropagation algorithm with the binary cross-entropy loss function \cite{geron2019hands}. To be more precise, the MLPNN structure attempts to learn the network parameter set $\boldsymbol{\Theta} = \{\boldsymbol{\Upsilon}, \boldsymbol{\xi}\}$ that basically contains the connection weight matrix $\boldsymbol{\Upsilon}$ and the bias vector $\boldsymbol{\xi}$. Under the multiclass classification framework, let $\mathbf{z}_n \in \{0,1\}^{N_c}$ be the $n$-th ground truth label vector with $N_c$ as the number of output classes, and consider $\mathbf{s}_n$ the corresponding input vector, therefore, the training set can be represented as $\Gamma = \{\mathbf{z}_n, \mathbf{s}_n\}_{n=1}^{N_t}$ with $N_t$ as the number of training samples. Hence, the training stage attempts learn the network parameters $\boldsymbol{\Theta}$ that minimize the loss function given by
\begin{equation}
    \mathcal{E}(\boldsymbol{\Theta}) = - \sum_{n=1}^{N_t} \sum_{c=1}^{N_c} (\mathbf{z}_n)_c \log \left(\left(\mathbf{p}_n(\boldsymbol{\Theta}, \mathbf{s}_n)\right)_c\right)
\end{equation}
where $(\mathbf{z}_n)_c$ is the $c$-th element of the ground truth label vector and $\left(\mathbf{p}_n(\boldsymbol{\Theta}, \mathbf{s}_n)\right)_c$ is the probability predicted by the network at the $c$-th output layer node. Notice that the network parameters are randomly initialized.




\vspace{-10pt}

\section{Results and analysis}
\label{sec:results}

\subsection{Pavia University}

\begin{figure}[h!]
\begin{center}
\begin{tabular}{cccc}
    \hspace{-10pt}
    \begin{subfigure}[b]{.14\textwidth}
    \begin{center}
        \includegraphics[width=\textwidth]{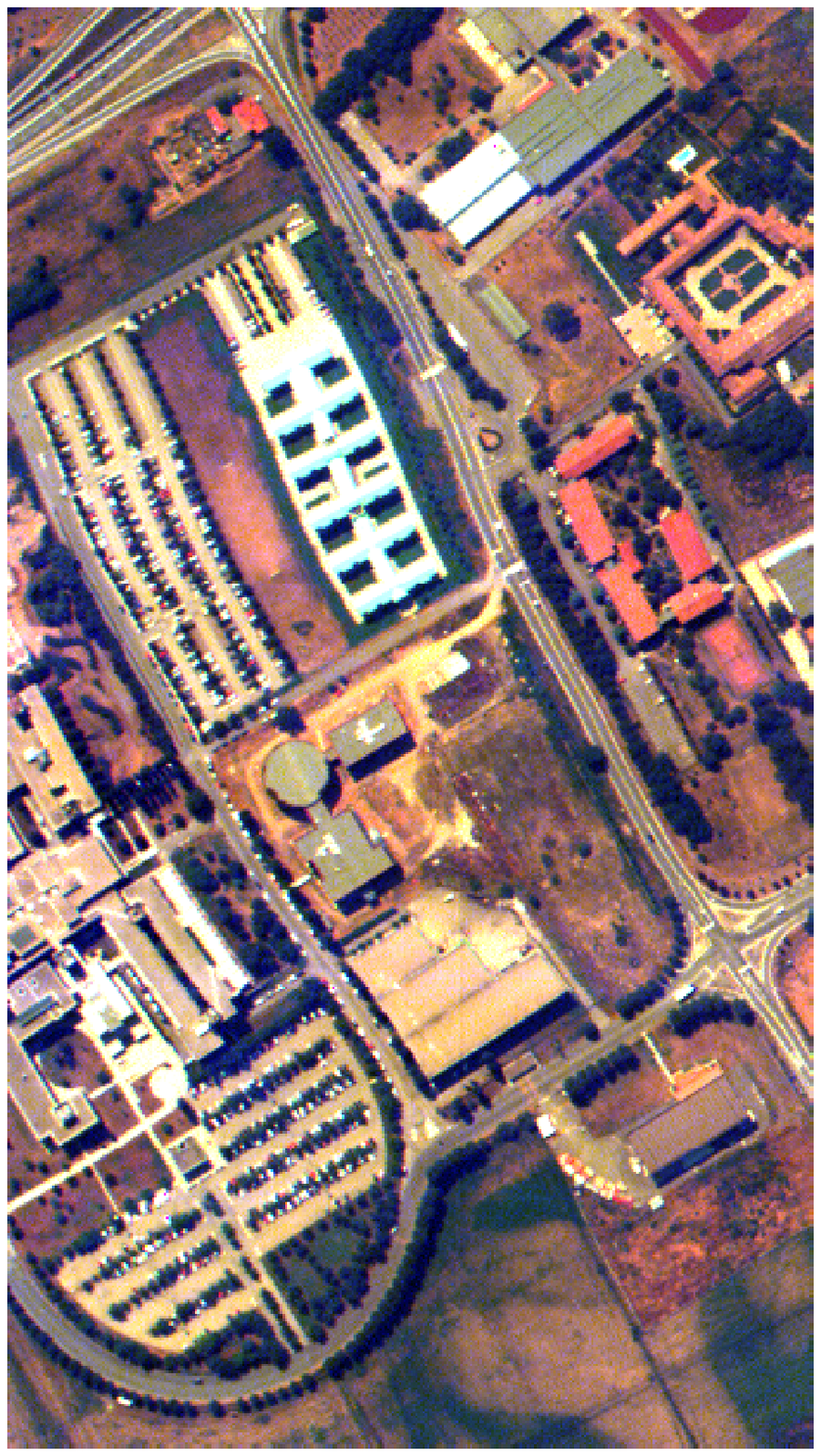}
    \end{center}
    \vspace{-15pt}
    \caption{}
    \label{subfig:rgb_hr_image}
    \end{subfigure}
    &
    \hspace{-10pt}
    \begin{subfigure}[b]{.14\textwidth}
    \begin{center}
        \includegraphics[width=\textwidth]{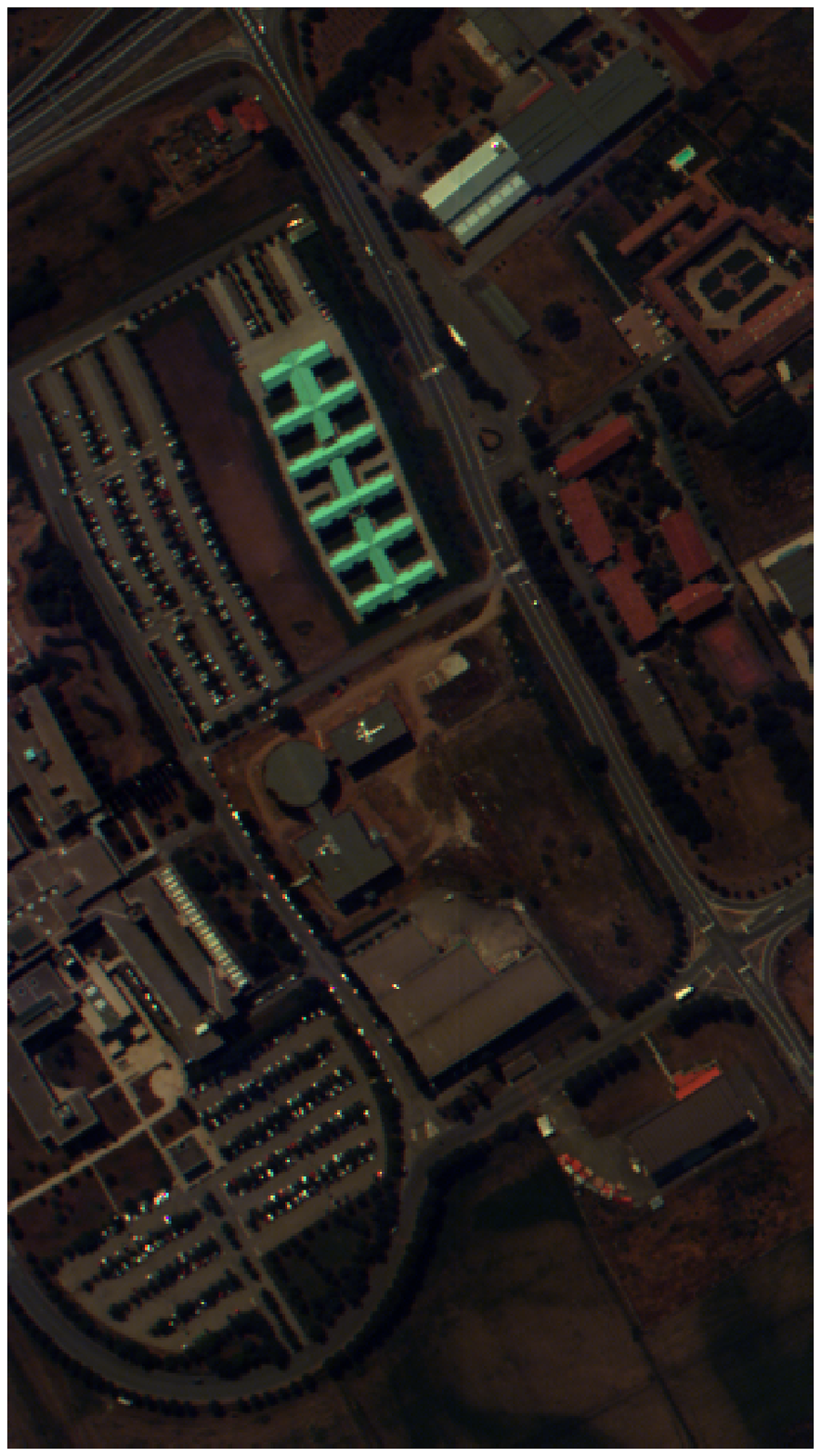}
    \end{center}
    \vspace{-15pt}
    \caption{}
    \label{subfig:rgb_ms_image}
    \end{subfigure}
    &
    \hspace{-10pt}
    \begin{subfigure}[b]{.14\textwidth}
    \begin{center}
        \includegraphics[width=\textwidth]{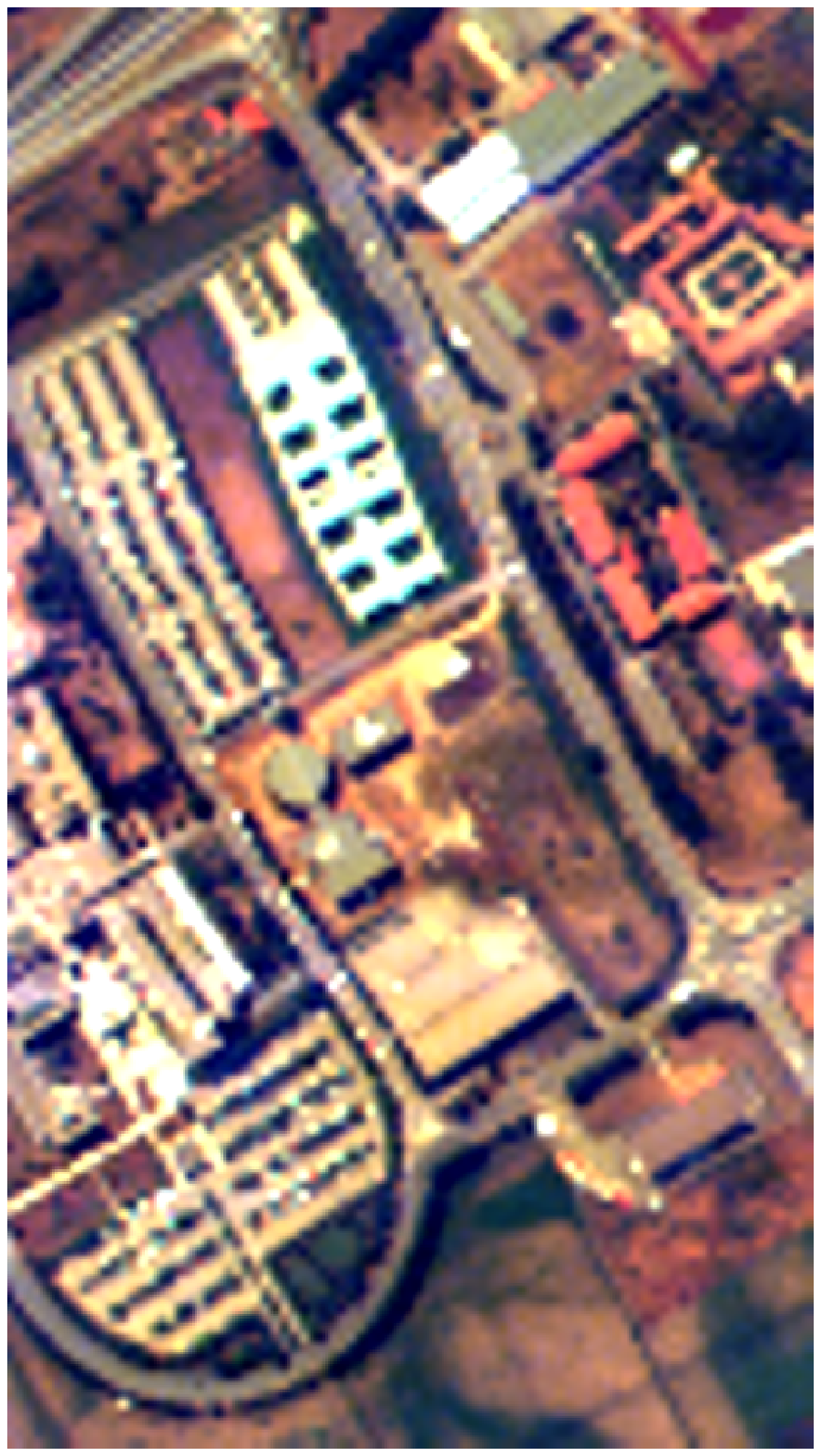}
    \end{center}
    \vspace{-15pt}
    \caption{}
    \label{subfig:rgb_hs_image}
    \end{subfigure}
    &
    \hspace{-10pt}
    \begin{subfigure}[b]{.14\textwidth}
    \begin{center}
        \includegraphics[width=\textwidth]{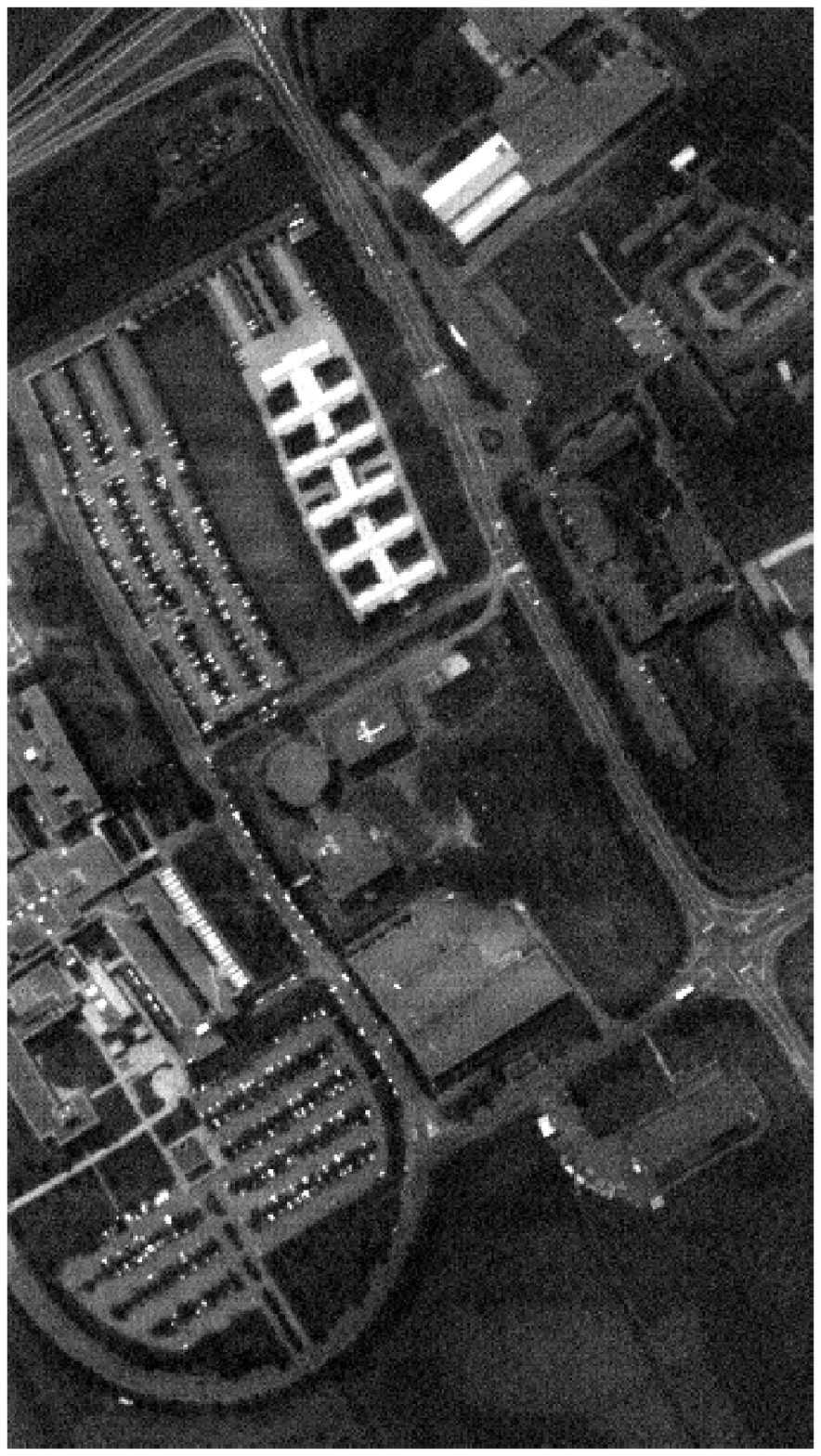}
    \end{center}
    \vspace{-15pt}
    \caption{}
    \label{subfig:rgb_hrf_band}
    \end{subfigure}
    \\
    \hspace{-10pt}
    \begin{subfigure}[b]{.14\textwidth}
    \begin{center}
        \includegraphics[width=\textwidth]{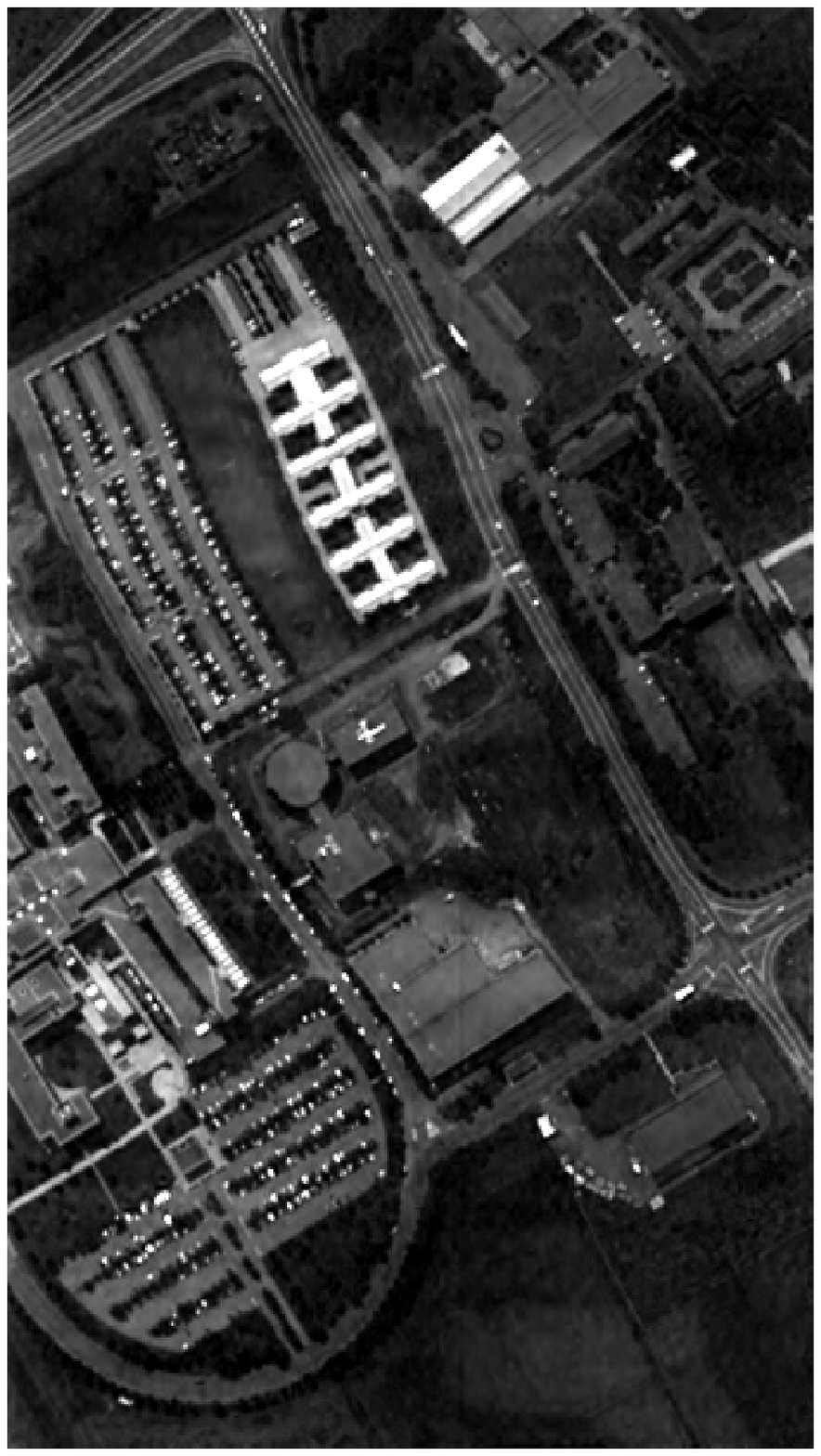}
    \end{center}
    \vspace{-15pt}
    \caption{$\lambda_2 = 0$}
    \label{subfig:feature_band1}
    \end{subfigure}
    &
    \hspace{-10pt}
    \begin{subfigure}[b]{.14\textwidth}
    \begin{center}
        \includegraphics[width=\textwidth]{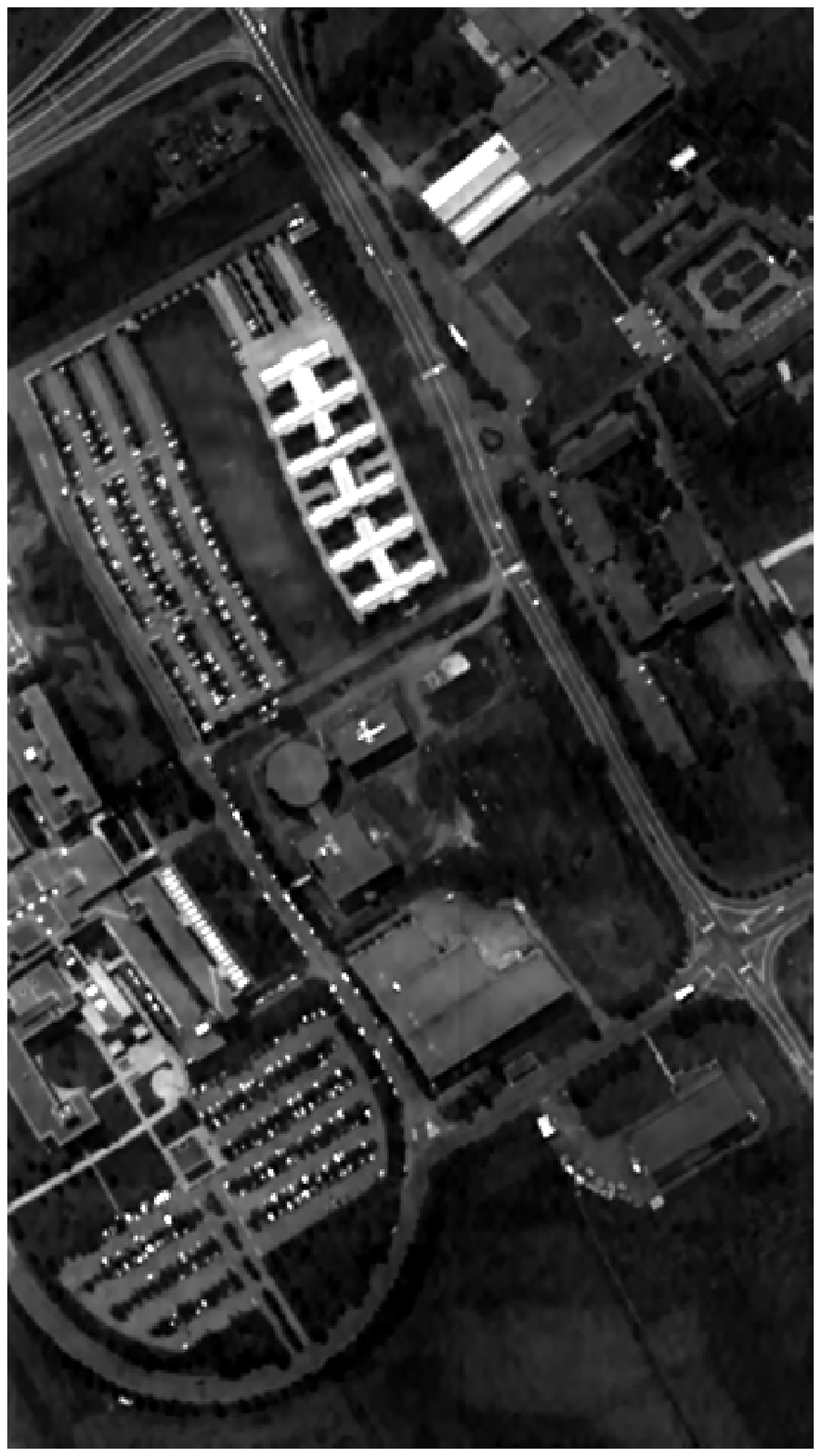}
    \end{center}
    \vspace{-15pt}
    \caption{$\lambda_2 = 1\times 10^{-4}$}
    \label{subfig:feature_band2}
    \end{subfigure}
    &
    \hspace{-10pt}
    \begin{subfigure}[b]{.14\textwidth}
    \begin{center}
        \includegraphics[width=\textwidth]{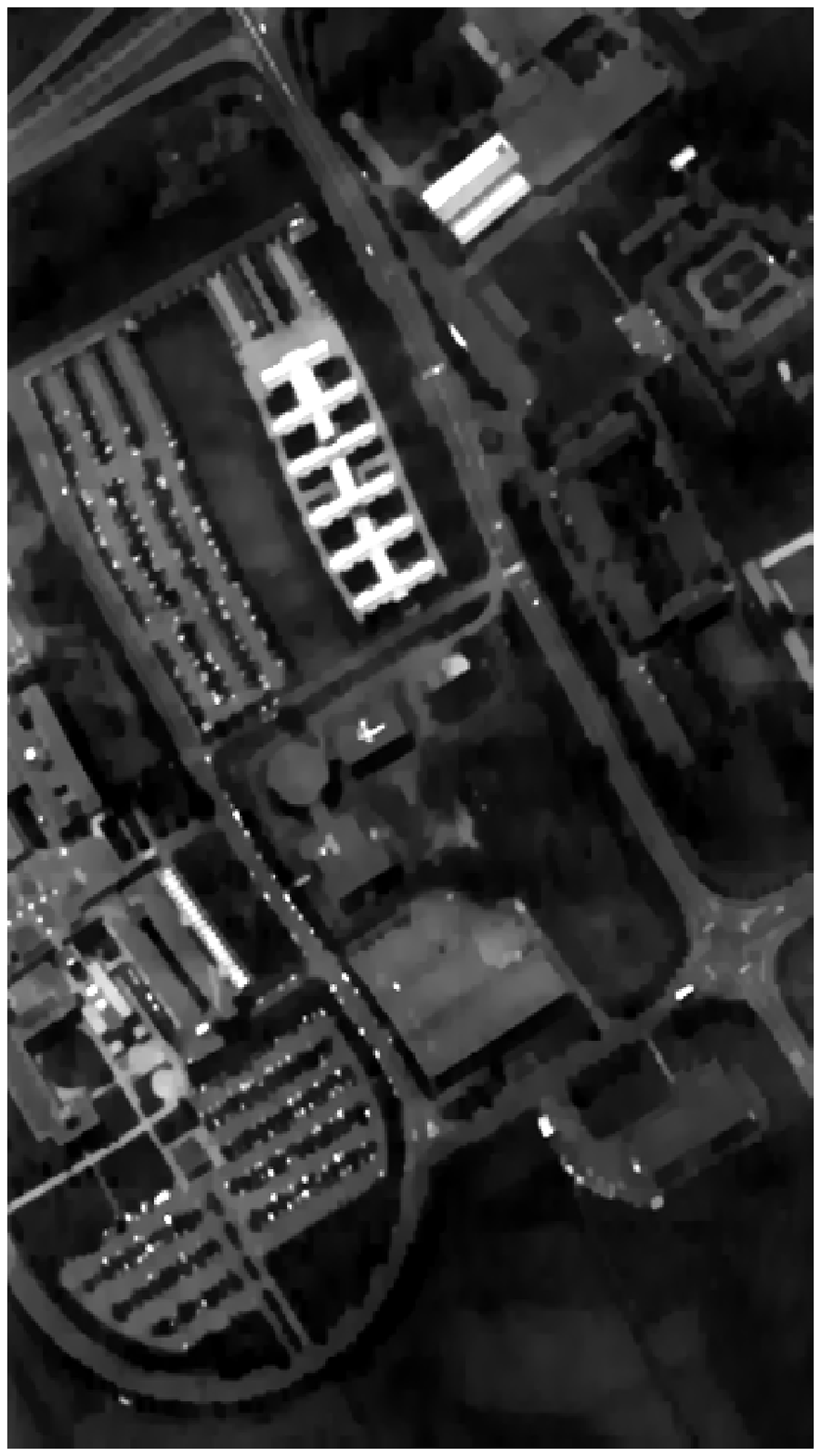}
    \end{center}
    \vspace{-15pt}
    \caption{$\lambda_2 = 2\times 10^{-4}$}
    \label{subfig:feature_band3}
    \end{subfigure}
    &
    \hspace{-10pt}
    \begin{subfigure}[b]{.14\textwidth}
    \begin{center}
        \includegraphics[width=\textwidth]{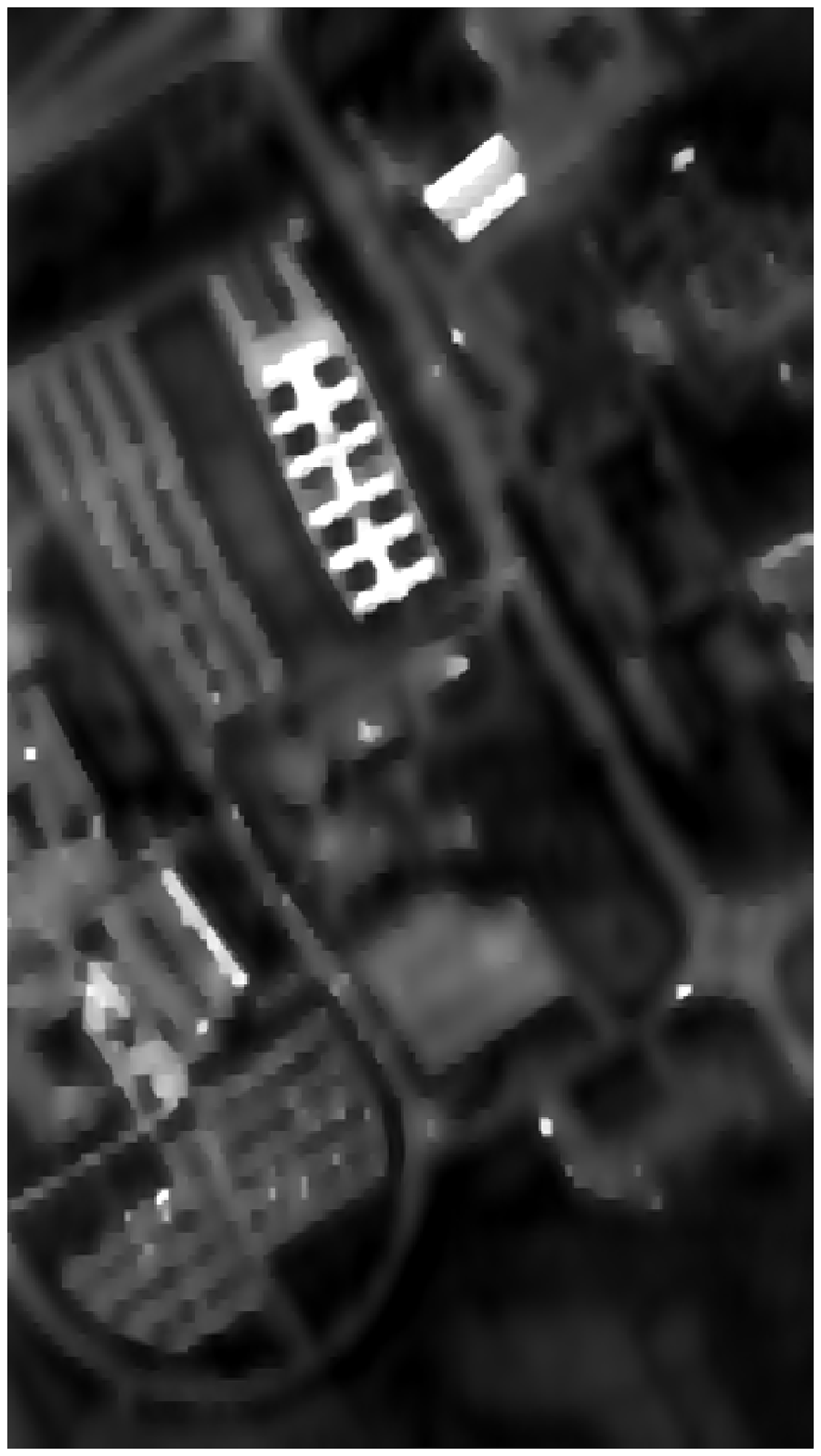}
    \end{center}
    \vspace{-15pt}
    \caption{$\lambda_2 = 5\times 10^{-4}$}
    \label{subfig:feature_band4}
    \end{subfigure}
\end{tabular}
\end{center}
    \vspace{-15pt}
    \caption{Pavia University spectral image. RGB composite of (a) the original image, (b) the multispectral image, and (c) the hyperspectral image; (d) reference feature band; (e)-(h) feature bands obtained by the proposed fusion method for different values the TV regularization parameter $\lambda_2$ with of $\lambda_1 = 0.001 \times \|\mathbf{H}^{\top}\mathbf{y}\|$.}
    \label{fig:my_label}
\end{figure}

This spectral image was acquired by the airborne sensor known as Reflective Optics Spectrographic Imaging System (ROSIS-03) over the University of Pavia, Italy \cite{SpectralImageDatabase}. This data set exhibits a high-spatial-resolution ($1.3~\mathrm{m}$ per pixel) with dimensions $610 \times 340$ pixels and $96$ spectral bands in the wavelength range from $0.43$ to $0.86$ $\mathrm{\mu m}$. Figure \ref{subfig:rgb_hr_image} displays the RGB composite of the Pavia University image.

Notice that the proposed fusion method is directly applied to compressive measurements of the MS image and the HS image. First, we built the MS and the HS versions of the input spectral image. To this end, the MS image was derived by integrating contiguous spectral bands of the original image with $q=4$. Therefore, the MS image is a high-spatial-resolution data cube with dimensions $610 \times 340 \times 24$. The RGB composite of the MS image is shown in Fig \ref{subfig:rgb_ms_image}. The HS image was obtained by applying spatial downsampling to every spectral band of the original image with $p=4$. The resulting image exhibits dimensions of $152 \times 85 \times 96$. Figure \ref{subfig:rgb_hs_image} shows the RGB composite of the HS image. For comparison purposes, Fig. display a band of the high-resolution features using the model described in (\ref{eq:hrf_model}).

Then, compressive measurements were obtained by simulating the 3D-CASSI dual-arm optical system. More precisely, the compressive measurements of the MS image comprise 6 high-spatial-resolution snapshots (compression ratio: $25\%$) while the compressive measurements of the HS image are 24 low-spatial-resolution projections (compression ratio: $25\%$). Subsequently, the proposed feature fusion method was applied directly to the compressive measurements. Figures \ref{subfig:feature_band1}-\ref{subfig:feature_band4} display the bands recovered by the proposed feature fusion algorithm for different values of the TV regularization parameter $\lambda_2$. For this experiment, the sparsity regularization parameter was set to $\lambda_1 = 0.001 \times \|\mathbf{H}^{\top} \mathbf{y} \|_{\infty}$. As can be seen in these figures, the recovered bands preserve the spatial structures of the underlying scene. It is relevant to notice that the fused cube is estimated directly from compressive measurements without applying a previous feature extraction stage. 

\begin{figure}
\begin{center}
\begin{tabular}{ccccc}
    \hspace{-10pt}
    \includegraphics[width=.15\textwidth]{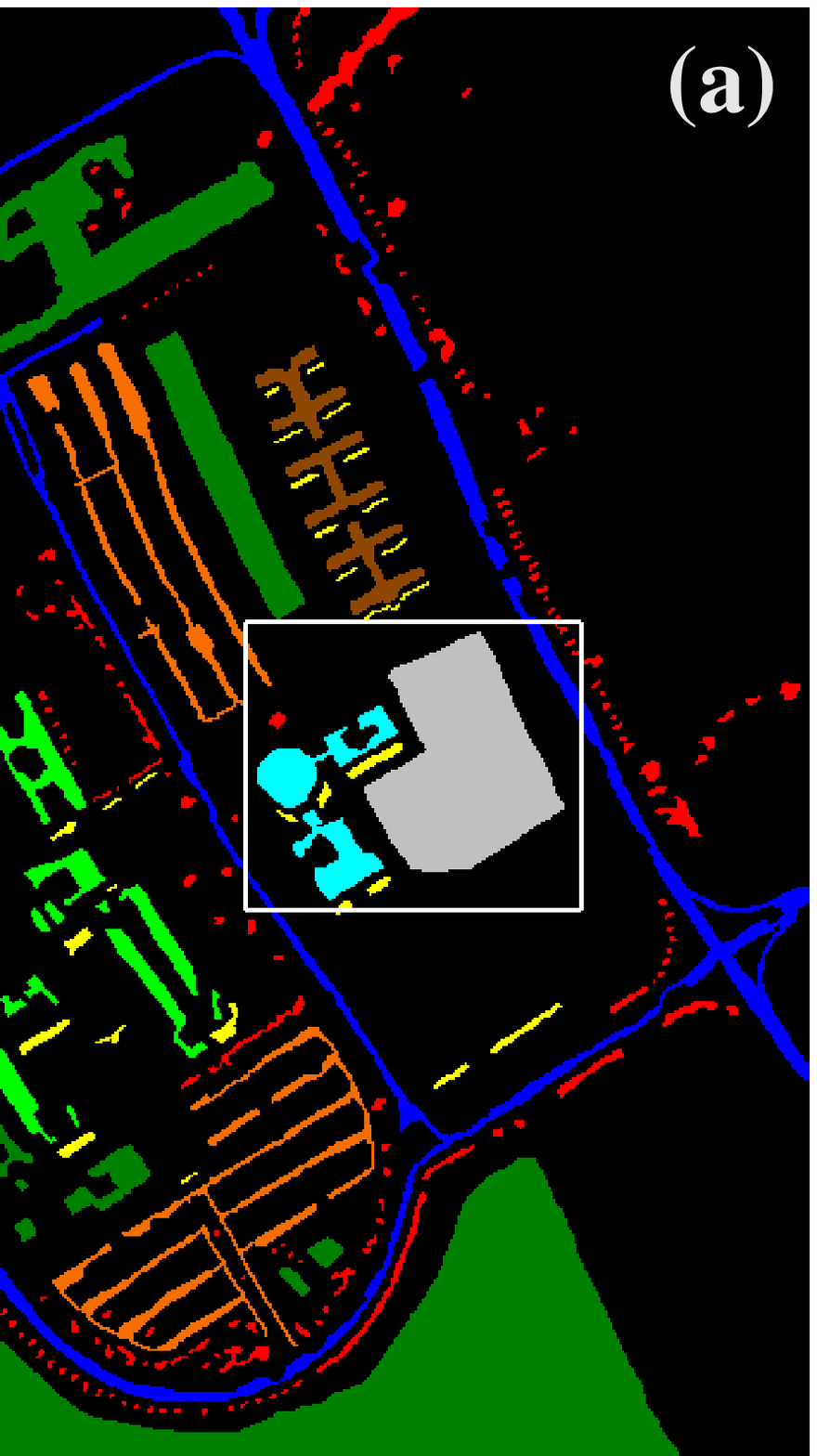}
     &
    \hspace{-15pt}
    \includegraphics[width=.15\textwidth]{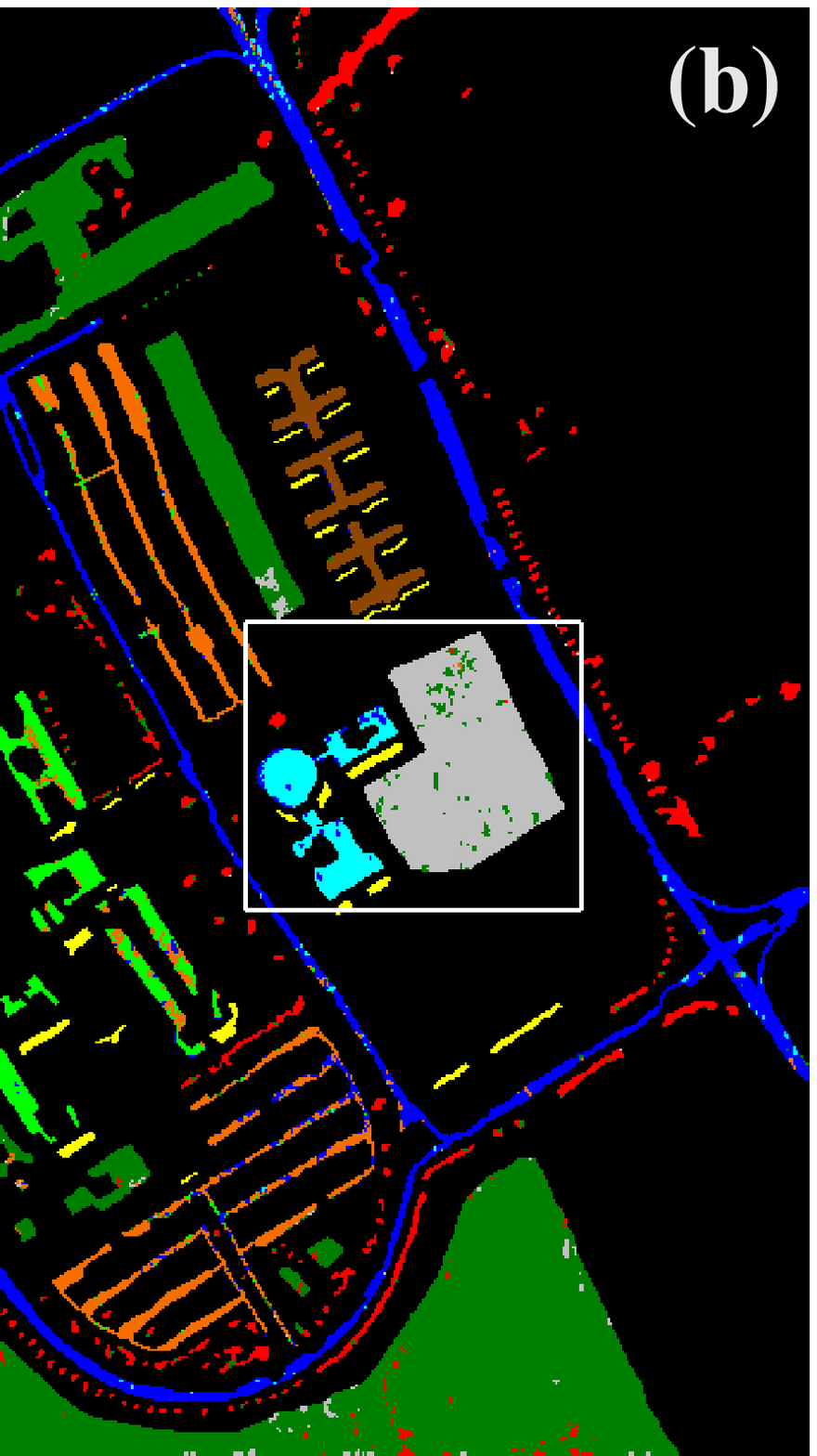}
     &
    \hspace{-15pt}
    \includegraphics[width=.15\textwidth]{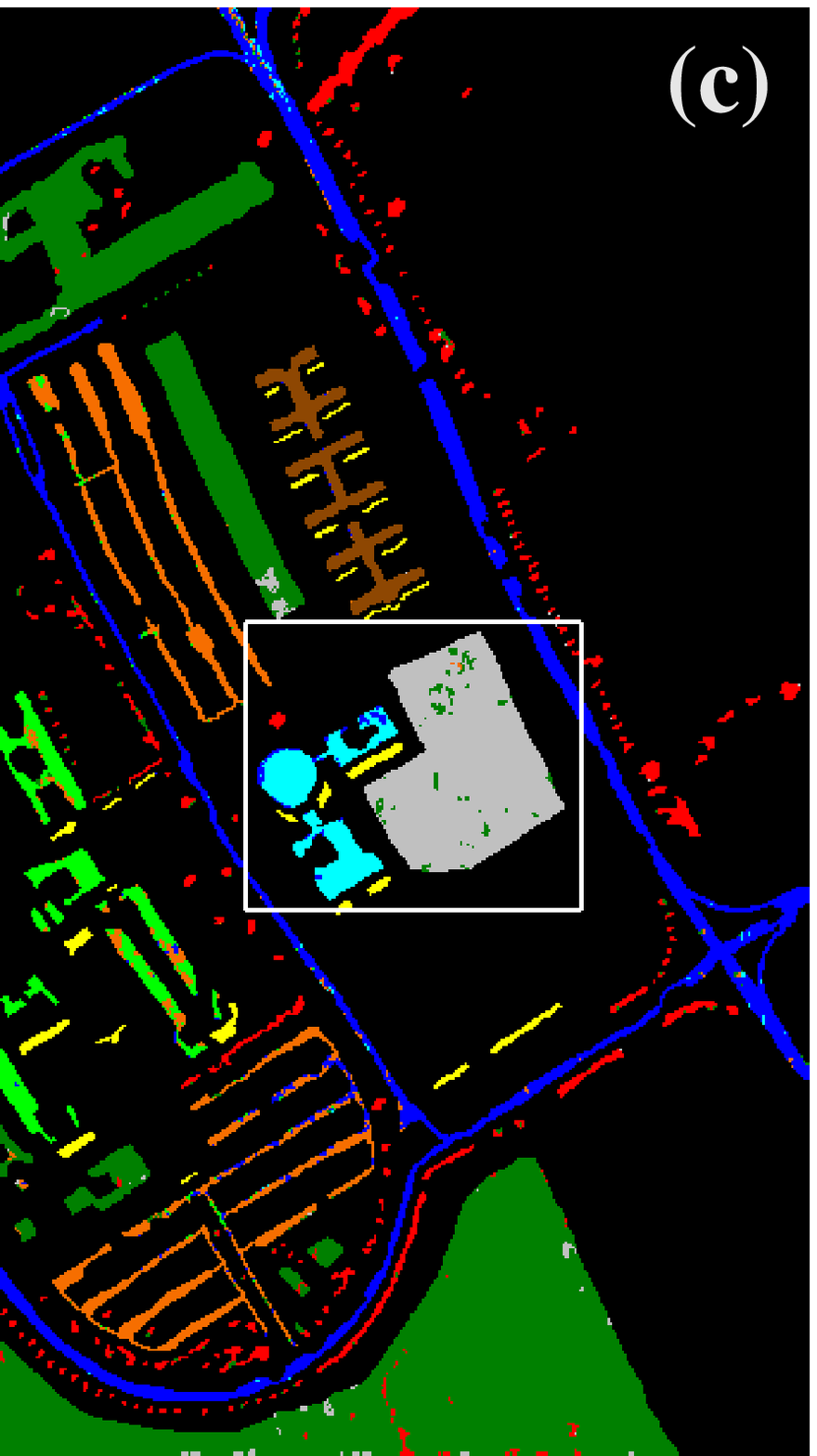}
     &
    \hspace{-15pt}
    \includegraphics[width=.15\textwidth]{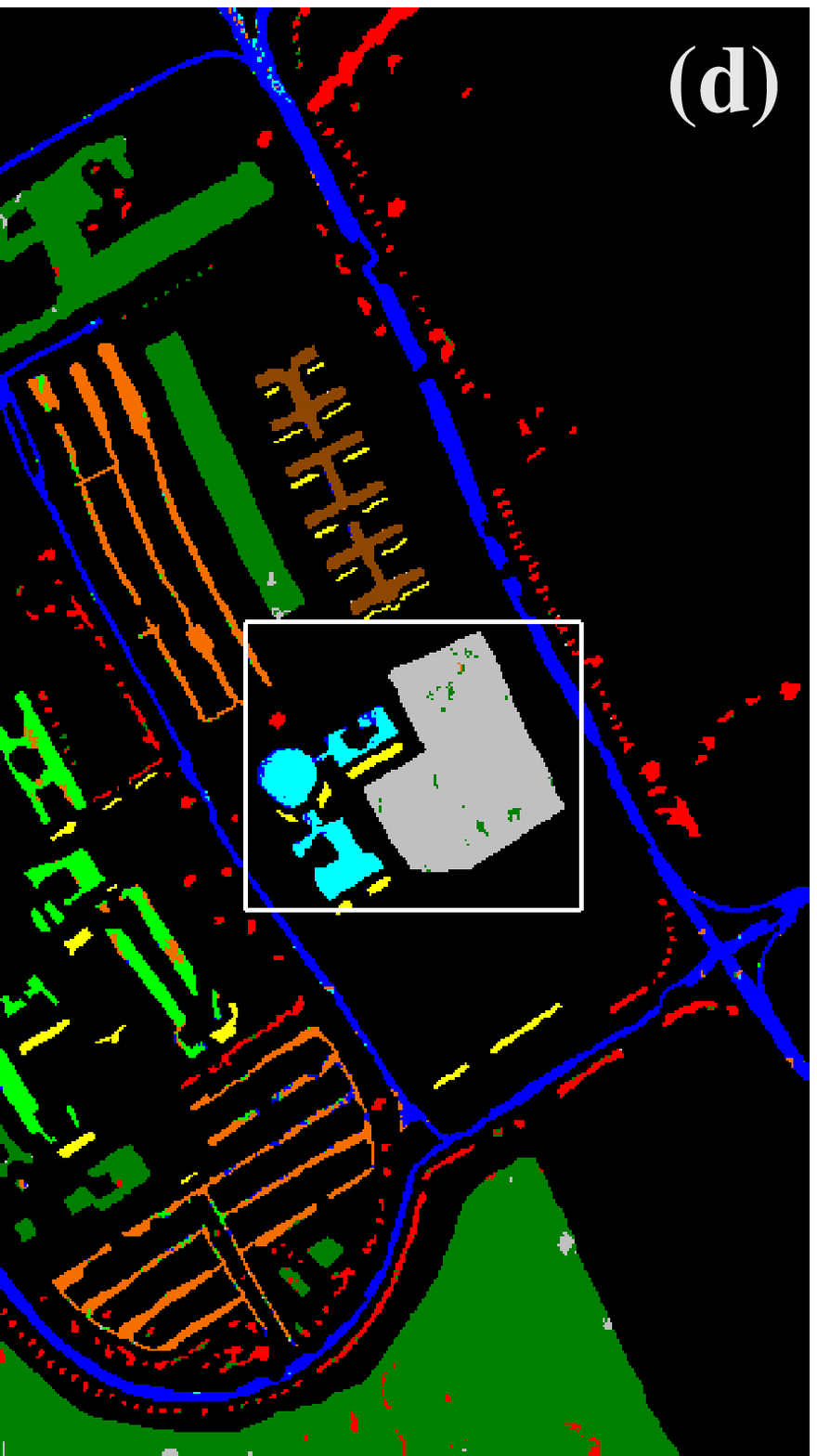}
    &
    \hspace{-15pt}
    \includegraphics[width=.15\textwidth]{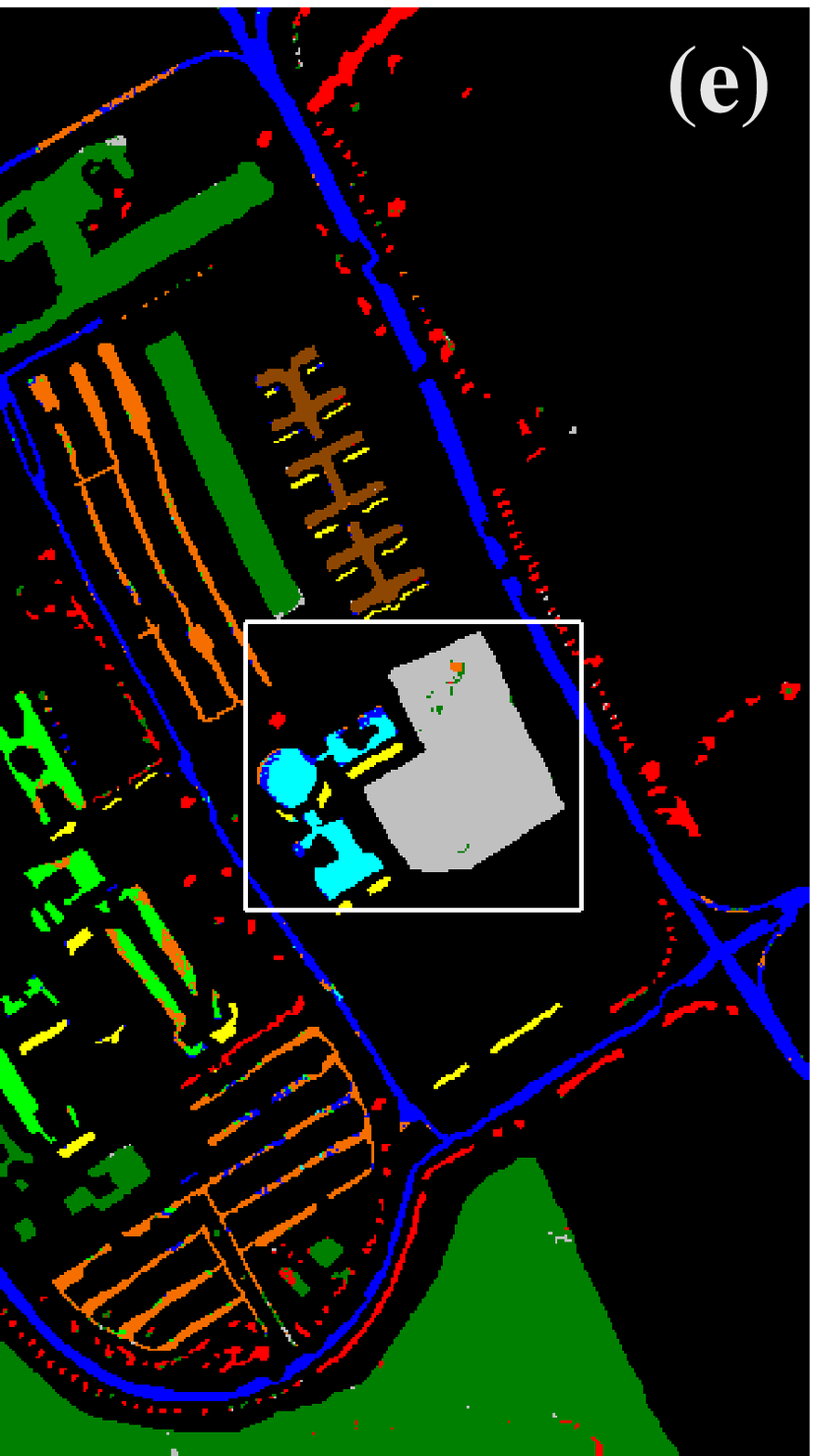}\vspace{-6pt}
    \\
    \hspace{-10pt}
    \includegraphics[width=.15\textwidth]{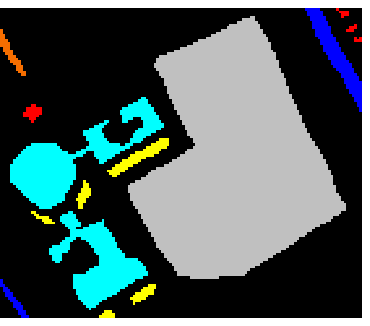}
     &
    \hspace{-15pt}
    \includegraphics[width=.15\textwidth]{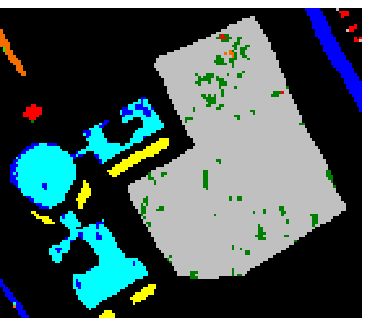}
     &
    \hspace{-15pt}
    \includegraphics[width=.15\textwidth]{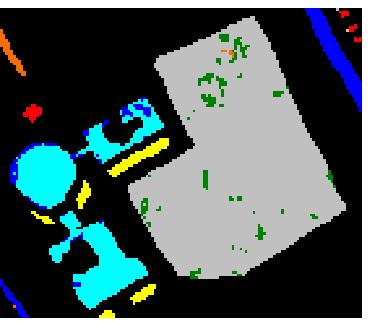}
     &
    \hspace{-15pt}
    \includegraphics[width=.15\textwidth]{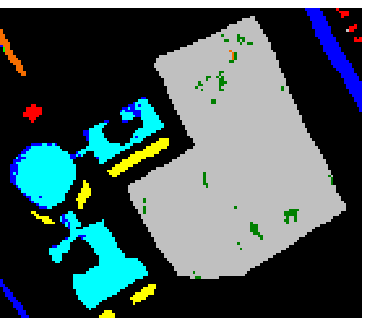}
    &
    \hspace{-15pt}
    \includegraphics[width=.15\textwidth]{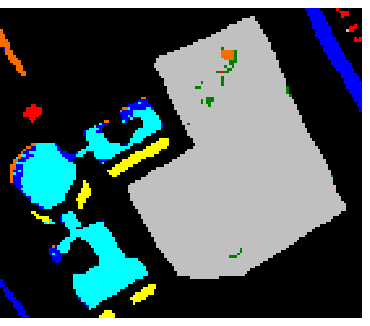}\vspace{10pt}
\end{tabular}
\end{center}
\vspace{-15pt}
    \caption{Pavia University spectral image. (a) The classification map of the ground truth; classification maps obtained by proposed approach for different TV regularization parameters (b) $\lambda_2 = 0$ (c) $\lambda_1 = 5 \times 10^{-6}$ (d) $\lambda_1 = 5 \times 10^{-5}$ (e) $\lambda_1 = 5 \times 10^{-4}$.}
    \vspace{-10pt}
    \label{fig:pavia_classmap}
\end{figure}

The bands of the fused features exhibit smoother regions as $\lambda_2$ increases, preserving, in turn, the spatial structure and the edges of the scene. Additionally, image noise is minimized as $\lambda_2$ increases, and this effect efficiently improve the labeling performance. To evaluate this behavior, the classification maps obtained for different values of the TV parameter are illustrated in Fig. \ref{fig:pavia_classmap}. The ground truth map is shown in Fig. \ref{fig:pavia_classmap}a. As can be seen in this figure, the classification noise is reduced as $\lambda_2$ increases.

\begin{table*}[]
    \ssmall
    \begin{center}
    \begin{tabular}{|c|c c|c c c c|c c c c|}
    \hline
    \hline
     & \multicolumn{2}{c|}{$\#$ Samples} & \multicolumn{4}{c|}{High-resolution spectral image} & \multicolumn{4}{c|}{Proposed feature fusion method} \\
     \hline
     Classes & Train & Test & SVM-RBF & RF & CNN & MLPNN & SVM-RBF & RF & CNN & MLPNN \\
    \hline
    \hline
    Asphalt & 661 & 5944 & 88.04 & \textbf{91.14} & 40.81 & 90.55 & 95.54 & 94.55 & 39.86 & \textbf{95.98} \\
    Meadows & 1865 & 16784 & \textbf{98.13} & 97.60 & 96.26 & 96.78 & 99.41 & 99.01 & 95.89 & \textbf{99.48} \\
    Gravel & 210 & 1889 & 65.08 & 68.68 & 19.34 & \textbf{72.92} & 69.18 & 74.42 & 20.01 & \textbf{80.22} \\
    Trees & 304 & 2736 & 70.26 & 88.94 & 5.54 & \textbf{90.95} & 95.43 & 93.35 & 8.98 & \textbf{95.61} \\
    Metal & 135 & 1210 & 59.73 & 98.66 & 43.74 & \textbf{99.32} & 99.08 & 99.15 & 58.41 & \textbf{99.42}\\
    Bare Soil & 503 & 4526 & 56.46 & 63.07 & \textbf{94.73} & 87.30 & 94.28 & 91.19 & 96.14 & \textbf{98.07} \\
    Bitumen & 133 & 1197 & 75.22 & 76.30 & 13.80 & \textbf{81.62} & 80.00 & 70.33 & 15.47 & \textbf{82.45} \\
    Bricks & 368 & 3314 & 84.82 & 87.53 & 12.61 & \textbf{89.04} & \textbf{90.30} & 87.30 & 20.31 & 86.69 \\
    Shadows & 95 & 852 & 92.69 & \textbf{99.77} & 21.95 & 99.39 & 96.85 & \textbf{97.92} & 19.39 & 96.58 \\
    \hline
    \hline
     \multicolumn{3}{|c|}{\multirow{2}{*}{Overall accuracy ($\%$)}} & 84.87 $\pm$ & 89.05 $\pm$ & 64.21 $\pm$ & \textbf{92.11 $\pm$} & 94.73 $\pm$ & 93.87 $\pm$ & 65.46 $\pm$ & \textbf{95.98 $\pm$} \\
     \multicolumn{3}{|c|}{} & 0.30 & 0.34 & 1.85 & \textbf{0.87} & 0.20 & 0.29 & 1.82 & \textbf{0.50} \\
     \hline
     \multicolumn{3}{|c|}{\multirow{2}{*}{Average accuracy ($\%$)}} & 76.71 $\pm$ & 85.74 $\pm$ & 38.76 $\pm$ & \textbf{89.76 $\pm$} & 90.23 $\pm$ & 89.69 $\pm$ & 41.61 $\pm$ & \textbf{92.72 $\pm$} \\
     \multicolumn{3}{|c|}{} & 0.68 & 0.57 & 2.74 & \textbf{1.19} & 0.40 & 0.41 & 2.06 & \textbf{0.79} \\
     \hline
     \multicolumn{3}{|c|}{\multirow{2}{*}{Kappa statistic}} & 0.790 $\pm$ & 0.852 $\pm$ & 0.491 $\pm$ & \textbf{0.895 $\pm$} & 0.930 $\pm$ & 0.918 $\pm$ & 0.516 $\pm$ & \textbf{0.945 $\pm$} \\
     \multicolumn{3}{|c|}{} & 0.005 & 0.005 & 0.028 & \textbf{0.012} & 0.002 & 0.004 & 0.021 & \textbf{0.008} \\
    \hline
    \hline
    \end{tabular}
    \end{center}
    \vspace{-15pt}
    \caption{Performance of the feature fusion method using different supervised classifiers on the Pavia University data set.}\vspace{-10pt}
    \label{tab:classifiers}
\end{table*}

To evaluate the performance of the proposed feature fusion method, Table \ref{tab:classifiers} shows the labeling accuracy results obtained by different supervised classification methods: support vector machines with radial basis function (SVM-RBF) kernel, random forest (RF), the convolutional neural network (CNN) and the multilayer perceptron neural network (MLPNN). To be more precise, the labeling accuracy is determined for each class, where each value is obtained by averaging ten realizations of the respective experiment. For each trial, a different pattern of the coded aperture is generated and a set of $10\%$ of features are randomly selected as training samples. For testing the corresponding classification method, the remaining $90\%$ of features are used. Furthermore, Table \ref{tab:classifiers} displays the overall accuracy (OA), the average accuracy (AA), and the Kappa statistic ($\kappa$) \cite{bishop2006machine}. For comparison purposes, the accuracy results obtained from the high-resolution image are also included, where spectral signatures are considered classification features. The best accuracy values obtained from either the high-resolution image and the fused features are in bold font. The fused features were estimated using $\lambda_1 = 0.01$ and $\lambda_2 = 5 \times 10^{-4}$.

For this experiment, the parameters of SVM-RBF classifier were set to $\sigma=1$ and $C=1$, where \textit{one-against-one} multi-classification rule was implemented \cite{MelganiClassification2004}. The number of trees of the RF classifier was fixed to 300. The CNN approach reduces the dimensionality of the input data via principal component analysis (PCA). Then, the first three principal components are used as input data of the CNN architecture. For each pixel, a patch of size $27 \times 27 \times 3$ is obtained with the aim of building the training and test sets. The CNN architecture consists of four convolutional layers of 16, 32, 64, and 128 filters, respectively, with a kernel size of $5 \times 5$. Furthermore, each hidden layer includes a max-pooling layer with a size of $2 \times 2$. Finally, a fully connected layer is aggregated at the output end. As can be seen in Table \ref{tab:classifiers}, the accuracy results obtained from fused features outperform those yielded from the high-resolution image. In addition, the results obtained by the MLPNN classifier are, in general, better than those yielded by the other labeling methods.

\subsection{Indian Pines}

\begin{table}[]
\ssmall
    \centering
    \begin{tabularx}{\linewidth}{|c|c|L|}
    \hline
    \hline
    \textbf{Method} & \textbf{Input}  & \textbf{Description}\\
    \hline
    \hline
    OTVCA \cite{RastiHyperspectral2016} & High-resolution spectral image & A feature extraction method for hyperspectral images that solves a TV penalized optimization problem. \\
    \hline
    SSLRA \cite{rasti2019hyperspectral} & High-resolution spectral image & A feature extraction method based on a low-rank model. This approach decomposes the hyperspectral image into smooth and sparse components. \\
    \hline
    EP \cite{GhamisiHyperspectral2016} & High-resolution spectral image & A feature extraction method based on the technique referred to as extended extinction profiles. \\
    \hline
    FES \cite{HinojosaSpectral2019} & Dual-resolution compressive measurements & Extracts features by reordering the compressive data yielding HS features and MS features. The method stacks an interpolated version of the HS features and a superpixel map of the MS features.\\
    \hline
    FEF-TR \cite{RamirezSpectral2020} & Dual-resolution compressive measurements & Implements the feature extraction stage \cite{HinojosaSpectral2019}. The extracted features are fused by solving an inverse problem penalized by the Tikhonov term that leads a closed-form solution. \\
    \hline
    FEF-L1TV \cite{RamirezMultiresolution2019} & Dual-resolution compressive measurements & Implements the feature extraction stage \cite{HinojosaSpectral2019}. The extracted features are fused by solving an inverse problem regularized by L1 and TV norms. An ADMM-based algorithm was developed to solve the formulated problem. \\
    \hline
    Proposed & Dual-resolution compressive measurements & Fuses features directly from compressive measurements by solving an inverse problem regularized by L1 and TV norms. An accelerated version of the ADMM algorithm is developed to solve the problem. \\
    \hline
    \hline
    \end{tabularx} \vspace{-10pt}
    \caption{General description of the comparison methods.}
    \label{tab:comparison}
\end{table}

\begin{figure}
\begin{center}
\begin{tabular}{cccc}
    \hspace{-10pt}
    \begin{subfigure}[b]{.24\textwidth}
    \begin{center}
        \includegraphics[width=\textwidth]{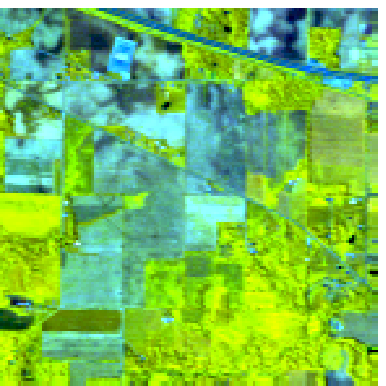}
    \end{center}
    \vspace{-15pt}
    \caption{}
    \label{subfig:rgb_hr_indian}
    \end{subfigure}
    &
    \hspace{-15pt}
    \begin{subfigure}[b]{.24\textwidth}
    \begin{center}
        \includegraphics[width=\textwidth]{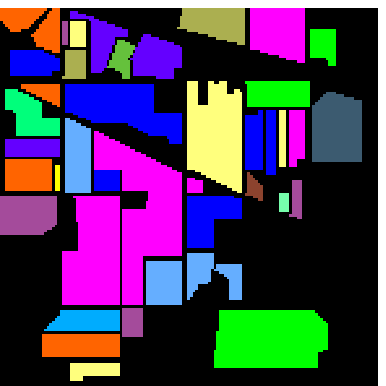}
    \end{center}
    \vspace{-15pt}
    \caption{}
    \label{subfig:gt_indian}
    \end{subfigure}
    &
    \hspace{-15pt}
    \begin{subfigure}[b]{.24\textwidth}
    \begin{center}
        \includegraphics[width=\textwidth]{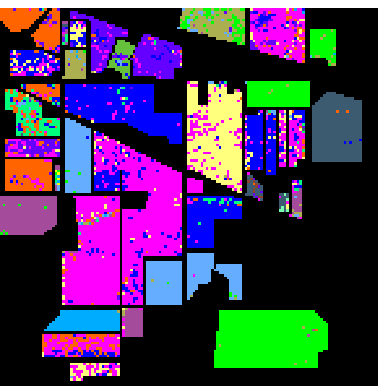}
    \end{center}
    \vspace{-15pt}
    \caption{}
    \label{subfig:cm_HR}
    \end{subfigure}
    &
    \hspace{-15pt}
    \begin{subfigure}[b]{.24\textwidth}
    \begin{center}
        \includegraphics[width=\textwidth]{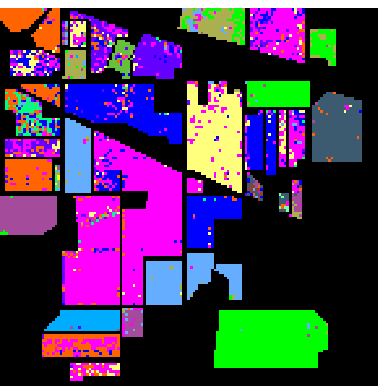}
    \end{center}
    \vspace{-15pt}
    \caption{}
    \label{subfig:cm_OTVCA}
    \end{subfigure}
    \\
    \hspace{-10pt}
    \begin{subfigure}[b]{.24\textwidth}
    \begin{center}
        \includegraphics[width=\textwidth]{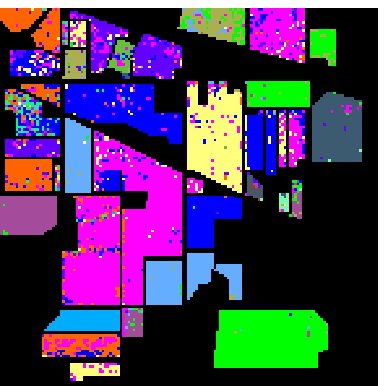}
    \end{center}
    \vspace{-15pt}
    \caption{}
    \label{subfig:cm_sslra}
    \end{subfigure}
    &
    \hspace{-15pt}
    \begin{subfigure}[b]{.24\textwidth}
    \begin{center}
        \includegraphics[width=\textwidth]{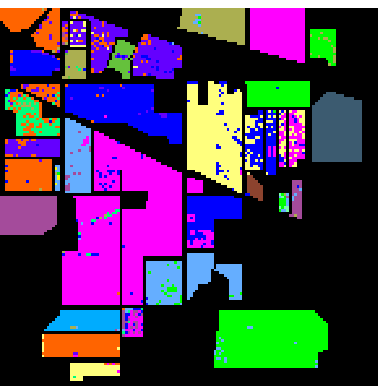}
    \end{center}
    \vspace{-15pt}
    \caption{}
    \label{subfig:cm_EP}
    \end{subfigure}
    &
    \hspace{-15pt}
    \begin{subfigure}[b]{.24\textwidth}
    \begin{center}
        \includegraphics[width=\textwidth]{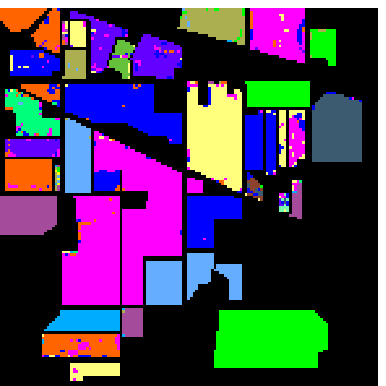}
    \end{center}
    \vspace{-15pt}
    \caption{}
    \label{subfig:cm_FEF}
    \end{subfigure}
    &
    \hspace{-15pt}
    \begin{subfigure}[b]{.24\textwidth}
    \begin{center}
        \includegraphics[width=\textwidth]{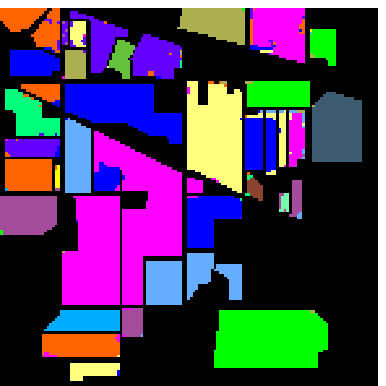}
    \end{center}
    \vspace{-15pt}
    \caption{}
    \label{subfig:cm_proposed}
    \end{subfigure}
\end{tabular}
\end{center}
    \vspace{-10pt}
    \caption{Indian Pines spectral image: (a) the RGB composite of the original image and (b) the classification map of the ground truth. Classification maps obtained by (c) the HR image (HR), OA: $79.40\%$, (d) the OTVCA method, OA: $79.47\%$, (e) the SSLRA method, OA: $81.31\%$, (f) the EP stacking, OA: $88.42\%$, (g) the FEF-L1TV method, OA: $92.99\%$, and (h) the proposed approach, OA: $96.54\%$. }
    \label{fig:my_label}
\end{figure}

\setlength{\tabcolsep}{4pt}
\begin{table*}[]
    \ssmall
    \begin{center}
    \begin{tabular}{|c|c c||c c c c c c|}
    \hline
    \hline
     \multirow{2}{*}{Classes} & \multicolumn{2}{c||}{$\#$ Samples} & \multirow{2}{*}{HR \cite{MelganiClassification2004}} & \multirow{2}{*}{OTVCA \cite{RastiHyperspectral2016}} & \multirow{2}{*}{SSLRA \cite{rasti2019hyperspectral}} & EP & \multirow{2}{*}{FEF-L1TV \cite{RamirezMultiresolution2019}} & \multirow{2}{*}{Proposed}  \\
     & Train & Test & & & & stacking \cite{GhamisiHyperspectral2016} & &  \\
    \hline
    \hline
    Alfalfa & 9 & 37 & $51.08\pm14.27$ & $18.54\pm9.60$ & $46.08\pm12.50$ & $68.35\pm8.31$ & $\mathbf{89.19}\pm\mathbf{8.55}$ & $86.22\pm8.96$ \\
    Corn-notill & 286 & 1142 & $71.18\pm3.38$ & $75.87\pm1.40$ & $74.59\pm2.53$ & $69.57\pm1.32$ & $95.58\pm1.26$ & $\mathbf{96.29}\pm \mathbf{1.26}$ \\
    Corn mintill & 166 & 664 & $54.86\pm2.56$ & $55.78\pm2.53$ & $57.24\pm2.36$ & $87.45\pm2.91$ & $95.32\pm1.63$ & $\mathbf{96.45}\pm\mathbf{1.25}$ \\
    Corn & 47 & 190 & $60.63\pm6.24$ & $32.76\pm4.78$ & $39.02\pm5.42$ & $54.52\pm11.10$ & $93.42\pm3.49$ & $\mathbf{95.21}\pm\mathbf{3.27}$ \\
    Grass-pasture & 97 & 386 & $86.76\pm2.51$ & $84.70\pm2.80$ & $88.89\pm2.19$ & $71.04\pm9.31$ & $\mathbf{94.61}\pm\mathbf{1.50}$ & $92.95\pm1.64$ \\
    Grass-trees & 146 & 584 & $96.23\pm0.97$ & $95.55\pm1.30$ & $97.68\pm0.95$ & $78.16\pm3.23$ & $99.26\pm0.31$ & $\mathbf{99.50}\pm\mathbf{0.40}$ \\
    Grass-pasture-mowed & 6 & 22 & $1.82\pm5.75$ & $49.86\pm11.97$ & $\mathbf{69.27}\pm\mathbf{12.43}$ & $44.77\pm11.78$ & $67.73\pm24.19$ & $42.73\pm29.00$ \\
    Hay-windrowed & 96 & 382 & $\mathbf{99.40}\pm\mathbf{0.35}$ & $98.96\pm0.73$ & $99.19\pm0.63$ & $99.23\pm0.38$ & $98.77\pm0.69$ & $98.95\pm0.76$ \\
    Oats & 4 & 16 & $0.00\pm0.00$ & $2.56\pm4.36$ & $3.06\pm5.65$ & $0.19\pm1.39$ & $\mathbf{65.00}\pm\mathbf{27.83}$ & $52.50\pm26.55$ \\
    Soybean-notill & 194 & 778 & $67.04\pm1.90$ & $70.55\pm2.53$ & $71.63\pm2.18$ & $87.61\pm1.93$ & $95.66\pm1.45$ & $\mathbf{96.75}\pm\mathbf{1.06}$ \\
    soybean-mintill & 491 & 1964 & $87.72\pm1.30$ & $86.11\pm1.15$ & $87.22\pm1.28$ & $95.04\pm1.35$ & $96.62\pm0.72$ & $\mathbf{97.56}\pm\mathbf{0.43}$ \\
    Soybean-clean & 119 & 474 & $69.11\pm2.86$ & $65.76\pm3.64$ & $70.54\pm4.70$ & $72.93\pm2.58$ & $92.09\pm2.37$ & \textbf{94.66 $\pm$ 2.76} \\
    Wheat & 41 & 164 & $93.90\pm3.76$ & $96.76\pm1.87$ & $95.04\pm2.53$ & $94.11\pm2.25$ & $\mathbf{97.99}\pm\mathbf{1.29}$ & $97.62\pm3.35$ \\
    Woods & 253 & 1012 & $97.80\pm0.28$ & $97.07\pm0.77$ & $98.12\pm0.64$ & $92.65\pm3.23$ & $\mathbf{99.25}\pm\mathbf{0.38}$ & $99.16\pm0.65$ \\
    Buildings-Drives & 77 & 309 & $54.95\pm2.97$ & $64.68\pm4.02$ & $75.06\pm3.92$ & $91.16\pm3.54$ & $96.28\pm2.16$ & $\mathbf{98.61}\pm\mathbf{1.40}$ \\
    Stone-Steel-Towers & 19 & 74 & $76.89\pm4.05$ & $69.03\pm6.68$ & $74.61\pm5.90$ & $81.77\pm6.81$ & $93.65\pm4.23$ & $\mathbf{94.86}\pm\mathbf{3.48}$ \\
    \hline
    \hline
    \multicolumn{3}{|c|}{Overall accuracy ($\%$)} & $79.66\pm0.56$ & $79.58\pm0.41$ & $81.38\pm0.54$ & $84.80\pm0.66$ & $96.27\pm0.24$ & $\mathbf{96.91}\pm\mathbf{0.38}$ \\
    \multicolumn{3}{|c|}{Average accuracy ($\%$)} & 66.84 $\pm$ 1.18 & $66.53\pm1.25$ & $71.70\pm1.37$ & $74.28\pm1.36$ & $\mathbf{91.90}\pm\mathbf{2.23}$ & $90.00\pm1.93$ \\
    \multicolumn{3}{|c|}{Kappa Statistic} & $0.766\pm0.007$ & $0.765\pm0.005$ & $0.786\pm0.006$ & $0.827\pm0.008$ & $0.965\pm0.003$ & $\mathbf{0.958}\pm\mathbf{0.004}$ \\
    \hline
    \hline
    \end{tabular}
    \end{center}
    \vspace{-10pt}
    \caption{Performance of various spectral image classification methods on the Indian Pines data set.}
    \label{tab:classifiers2}
\end{table*}

This data set was captured by an Airborne Visible/Infrared Imaging Spectrometer (AVIRIS) sensor over an agricultural region in Indiana, USA. This Pines spectral image consists of a dataset with $145 \times 145$ pixels and 200 spectral bands recording the wavelength range from $0.40$ to $2.50$ $\mu$m \cite{IndianPinesDataset}. Furthermore, this data set exhibits a spatial-resolution of $20$ m per pixel. The RGB composite of this spectral image is shown in Fig. \ref{subfig:rgb_hr_indian} and the ground truth map is displayed in Fig. \ref{subfig:gt_indian}.

For comparative purposes, Figs \ref{subfig:cm_HR}-\ref{subfig:cm_proposed} display the classification maps yielded by different types of features extraction and fusion methods. In particular, these methods obtain the classification features from the original spectral image. Specifically, we classify the spectral image from the spectral signatures of the high-resolution image (HR) \cite{MelganiClassification2004}, and the features obtained by the orthogonal total variation component analysis (OTVCA) \cite{RastiHyperspectral2016}, the sparse-smooth low-rank analysis (SSLRA) \cite{rasti2019hyperspectral}, and the extinction profile (EP) method \cite{GhamisiHyperspectral2016}. Additionally, Fig. \ref{subfig:cm_FEF} includes the labeling map obtained by the feature extraction and fusion technique with the L1-TV regularization (FEF-L1TV) \cite{RamirezMultiresolution2019} and the classification map yielded by the proposed classification approach is illustrated in Fig. \ref{subfig:cm_proposed}. Table \ref{tab:comparison} include a general description of the involved methods. For these figures, simulations selected $20\%$ of the classification attributes as training samples and $80\%$ of the remaining features to test the various labeling approaches. Furthermore, we use the same training and test sets to evaluate the various approaches. As can be seen, the proposed approach is not affected by the classification noise compared with respect to the other approaches, yielding homogeneous labeling regions. Notice that the labeling map obtained by the proposed classification method provides the best OA value.

Table \ref{tab:classifiers2} shows the accuracy results obtained by various classification techniques on the labeling of the Indian Pines dataset. The accuracy values are obtained by averaging ten realizations of the corresponding experiment. Note that the best accuracy results are in bold font. Furthermore, the overall accuracy (OA), the average accuracy (AA), and the Kappa Statistic ($\kappa$) are included in the last three rows of Table \ref{tab:classifiers2}. As can be observed in this table, the classification approach that contains the proposed feature fusion method exhibits, in general, the best performance.

\subsection{Salinas Valley}
\label{subsec:salinas}

\begin{figure}
\begin{center}
\begin{tabular}{ccc}
    \hspace{-15pt}
    \begin{subfigure}[b]{.15\textwidth}
    \begin{center}
        \includegraphics[width=0.8\textwidth]{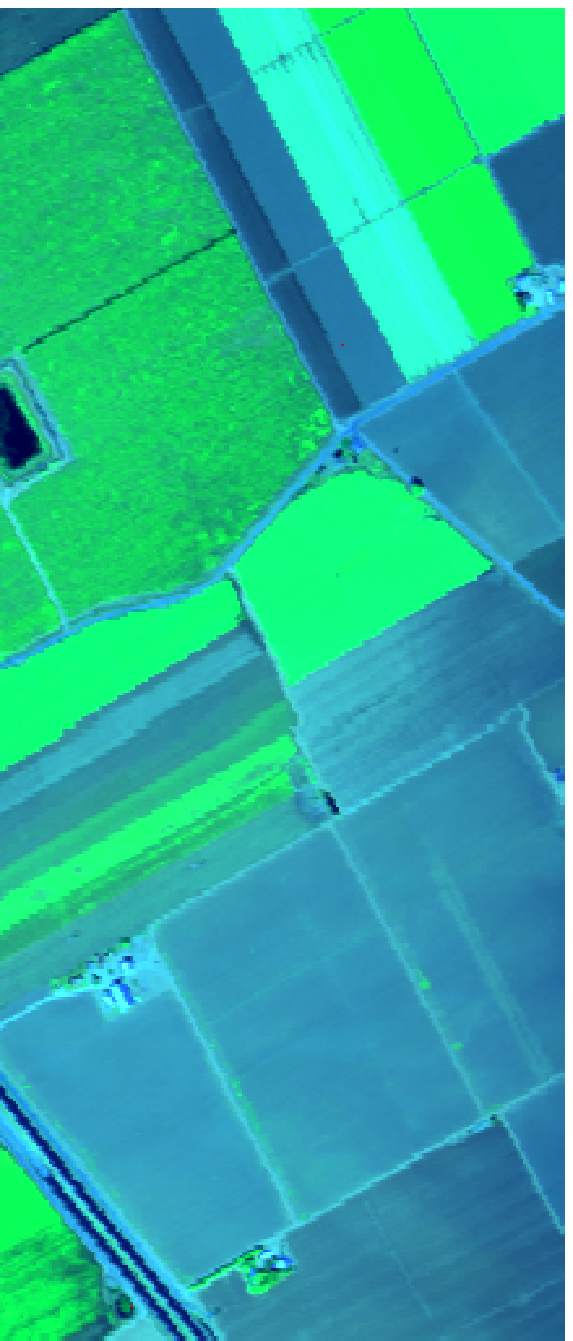}
    \end{center}
    \vspace{-15pt}
    \caption{}
    \label{subfig:rgb_hr_salinas}
    \end{subfigure}
    &
    \hspace{-15pt}
    \begin{subfigure}[b]{.12\textwidth}
    \begin{center}
        \includegraphics[width=\textwidth]{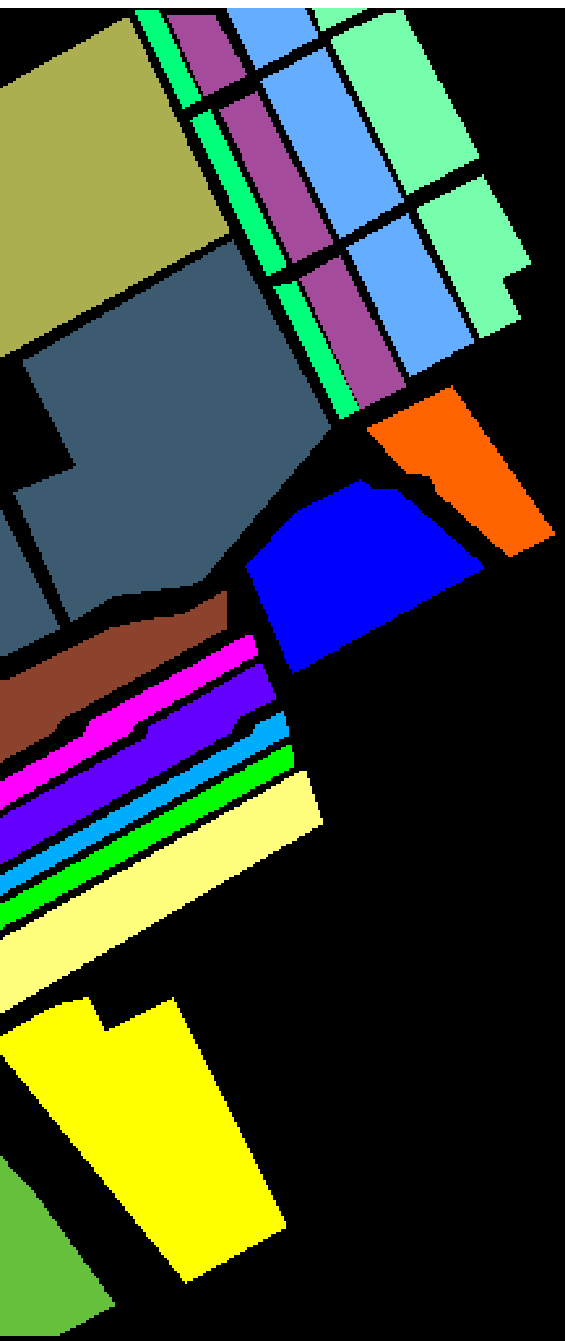}
    \end{center}
    \vspace{-15pt}
    \caption{}
    \label{subfig:gt_salinas}
    \end{subfigure}
    &
    \hspace{-15pt}
    \begin{subfigure}[b]{.12\textwidth}
    \begin{center}
        \includegraphics[width=1.45\textwidth]{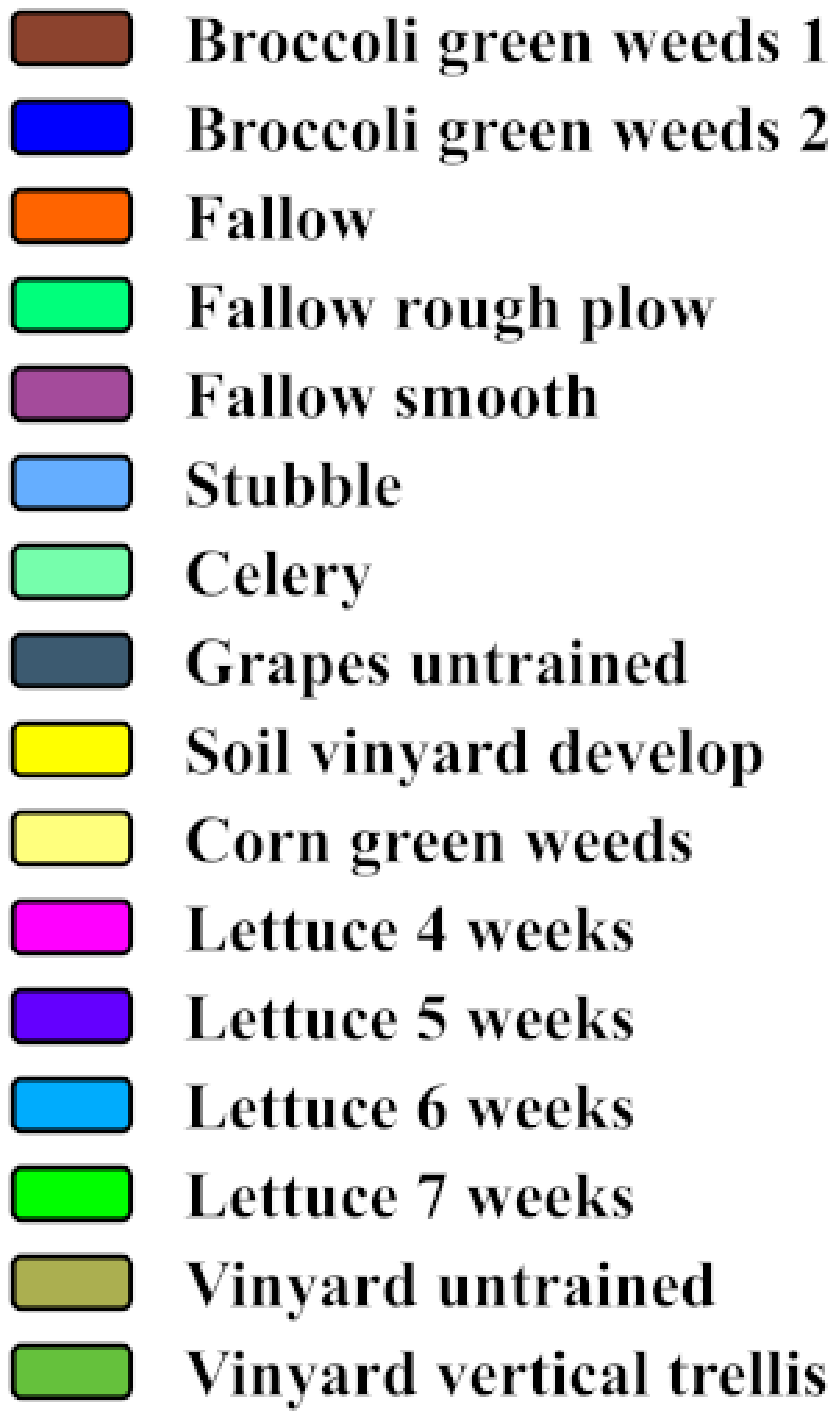}
    \end{center}
    \vspace{-15pt}
    \caption{}
    \label{subfig:classes_salinas}
    \end{subfigure}
\end{tabular}
\end{center}
    \vspace{-10pt}
    \caption{Salinas Valley spectral image: (a) the RGB composite of the original image and (b) the classification map of the ground truth.}
    \label{fig:my_label}
\end{figure}

The Salinas Valley spectral image was captured by an AVIRIS sensor over the Valley of Salinas, California, USA \cite{SalinasValleyDataset}. This spectral image has dimensions $512 \times 217 \times 192$ in the wavelength interval from $0.24$ to $2.40$ $\mu m$. The RGB composite of this image is shown in Fig. \ref{subfig:rgb_hr_salinas}. Furthermore, Figs \ref{subfig:gt_salinas} and \ref{subfig:classes_salinas} display the ground truth labeling map for $16$ different classes, where each class corresponds to a particular kind of crop.

\begin{figure}
\begin{center}
\begin{tabular}{c c c c}
    \hspace{-55pt}
    \begin{subfigure}[b]{.40\textwidth}
    \begin{center}
    	\begin{tikzpicture}
		\begin{axis}[
		legend cell align = left,
		legend pos = south east,
		scale = 0.45,
		grid = major,
		ylabel = \footnotesize \textrm{Overall accuracy ($\%$)},
		ylabel style = {yshift=-0.2cm},
		ymin = 80,
		ymax = 100,
		xmin = 12.5,
		xmax = 75,
		xlabel = \footnotesize \textrm{Compression ratio ($\%$)},
		ticklabel style = {font=\scriptsize},
		]
		\addplot[color = red, smooth, thick, line width=2pt] table[x ={x},y = y4]{FigureData/OAvsCompressionSalinas512.dat};
		\addlegendentry{\tiny \textrm{FES}}
		\addplot[color=green, smooth, thick, line width=2pt] table[x ={x},y = y2]{FigureData/OAvsCompressionSalinas512.dat};
		\addlegendentry{\tiny \textrm{FEF-TR}}
		\addplot[color=orange, smooth, thick, line width=2pt] table[x ={x},y = y3]{FigureData/OAvsCompressionSalinas512.dat};
		\addlegendentry{\tiny \textrm{FEF-L1TV}}
		\addplot[color=blue, smooth, thick, line width=2pt] table[x ={x},y = y1]{FigureData/OAvsCompressionSalinas512.dat};
		\addlegendentry{\tiny \textrm{Proposed}}
		\end{axis}
	\end{tikzpicture}
    
    \end{center}
    \vspace{-15pt}
    \caption{}
    \label{subfig:OAvsComp}
    \end{subfigure}
    &
    \hspace{-75pt}
        \begin{subfigure}[b]{.40\textwidth}
    \begin{center}
    	\begin{tikzpicture}
		\begin{axis}[
		legend cell align = left,
		legend pos = south east,
		scale = 0.45,
		grid = major,
		ylabel = \footnotesize \textrm{Overall accuracy ($\%$)},
		ylabel style = {yshift=-0.2cm},
		ymin = 80,
		ymax = 100,
		xmin = 1,
		xmax = 20,
		xlabel = \footnotesize \textrm{Training rate ($\%$)},
		ticklabel style = {font=\scriptsize},
		]
		\addplot[color = red, smooth, thick, line width=2pt] table[x ={x},y = y4]{FigureData/OAvsTrainingSalinas512.dat};
		\addlegendentry{\tiny \textrm{FES}}
		\addplot[color=green, smooth, thick, line width=2pt] table[x ={x},y = y2]{FigureData/OAvsTrainingSalinas512.dat};
		\addlegendentry{\tiny \textrm{FEF-TR}}
		\addplot[color=orange, smooth, thick, line width=2pt] table[x ={x},y = y3]{FigureData/OAvsTrainingSalinas512.dat};
		\addlegendentry{\tiny \textrm{FEF-L1TV}}
		\addplot[color=blue, smooth, thick, line width=2pt] table[x ={x},y = y1]{FigureData/OAvsTrainingSalinas512.dat};
		\addlegendentry{\tiny \textrm{Proposed}}
		\end{axis}
	\end{tikzpicture}
    
    \end{center}
    \vspace{-15pt}
    \caption{}
    \label{subfig:OAvsTraining}
    \end{subfigure}
    &
    \hspace{-75pt}
        \begin{subfigure}[b]{.40\textwidth}
    \begin{center}
    	\begin{tikzpicture}
		\begin{axis}[
		legend cell align = left,
		legend pos = south east,
		scale = 0.45,
		grid = major,
		ylabel = \footnotesize \textrm{Overall accuracy ($\%$)},
		ylabel style = {yshift=-0.2cm},
		ymin = 50,
		ymax = 100,
		xmin = 10,
		xmax = 20,
		xlabel = \footnotesize \textrm{SNR [dB]},
		ticklabel style = {font=\scriptsize},
		]
		\addplot[color = red, smooth, thick, line width=2pt] table[x ={x},y = y4]{FigureData/OAvsSNRGaussian.dat};
		\addlegendentry{\tiny \textrm{FES}}
		\addplot[color=green, smooth, thick, line width=2pt] table[x ={x},y = y2]{FigureData/OAvsSNRGaussian.dat};
		\addlegendentry{\tiny \textrm{FEF-TR}}
		\addplot[color=orange, smooth, thick, line width=2pt] table[x ={x},y = y3]{FigureData/OAvsSNRGaussian.dat};
		\addlegendentry{\tiny \textrm{FEF-L1TV}}
		\addplot[color=blue, smooth, thick, line width=2pt] table[x ={x},y = y1]{FigureData/OAvsSNRGaussian.dat};
		\addlegendentry{\tiny \textrm{Proposed}}
		\end{axis}
	\end{tikzpicture}
    
    \end{center}
    \vspace{-15pt}
    \caption{}
    \label{subfig:OAvsSNRG}
    \end{subfigure}
    &
    \hspace{-75pt}
        \begin{subfigure}[b]{.40\textwidth}
    \begin{center}
    	\begin{tikzpicture}
		\begin{axis}[
		legend cell align = left,
		legend pos = south east,
		scale = 0.45,
		grid = major,
		ylabel = \footnotesize \textrm{Overall accuracy ($\%$)},
		ylabel style = {yshift=-0.2cm},
		ymin = 50,
		ymax = 100,
		xmin = 10,
		xmax = 20,
		xlabel = \footnotesize \textrm{SNR [dB]},
		ticklabel style = {font=\scriptsize},
		]
		\addplot[color = red, smooth, thick, line width=2pt] table[x ={x},y = y4]{FigureData/OAvsSNRPoisson.dat};
		\addlegendentry{\tiny \textrm{FES}}
		\addplot[color=green, smooth, thick, line width=2pt] table[x ={x},y = y2]{FigureData/OAvsSNRPoisson.dat};
		\addlegendentry{\tiny \textrm{FEF-TR}}
		\addplot[color=orange, smooth, thick, line width=2pt] table[x ={x},y = y3]{FigureData/OAvsSNRPoisson.dat};
		\addlegendentry{\tiny \textrm{FEF-L1TV}}
		\addplot[color=blue, smooth, thick, line width=2pt] table[x ={x},y = y1]{FigureData/OAvsSNRPoisson.dat};
		\addlegendentry{\tiny \textrm{Proposed}}
		\end{axis}
	\end{tikzpicture}
    
    \end{center}
    \vspace{-15pt}
    \caption{}
    \label{subfig:OAvsSNRP}
    \end{subfigure}
\end{tabular}
\end{center}
\vspace{-25pt}
    \caption{Performance on the Salinas Valley data set of the various classification methods: (a) OA versus the compression ratio and (b) OA versus the rate of training samples. OA on the Salinas Valley spectral image yielded by the various classification approaches as the SNR increases for (c) Gaussian noise and (d) Poisson noise.}
\end{figure}
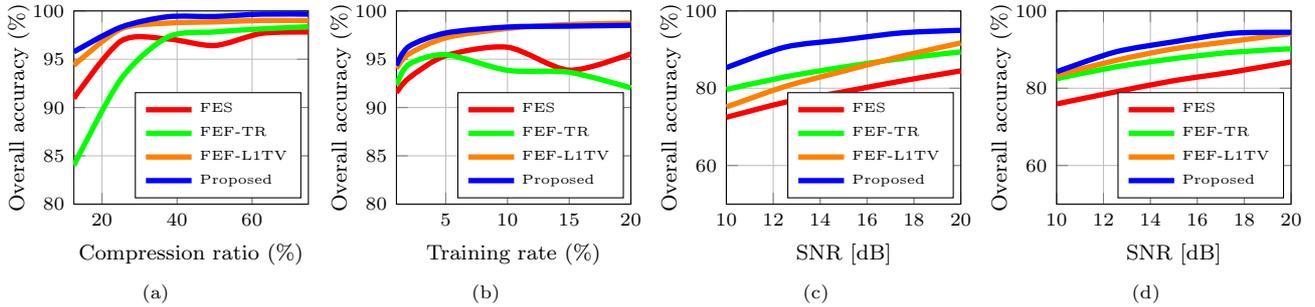

Figure \ref{subfig:OAvsComp} shows the overall accuracy curves as the compression ratio increases for different fusion methods that obtain features from dual-resolution compressive measurements. We compare the performance of the proposed method with respect to those yielded by the feature extraction and stacking method (FES) \cite{HinojosaSpectral2019}, the feature extraction and fusion technique that uses the Tikhonov regularization (FEF-TR) \cite{RamirezSpectral2020}, and the feature extraction and fusion method that exploits the L1-TV regularization (FEF-L1TV) \cite{RamirezMultiresolution2019}. In this work, the information embedded in coded aperture patterns is included in the projection matrices (\ref{eq:pmatrix_ms}) and (\ref{eq:pmatrix_hs}) leading to an algorithm that fuses spectral image features directly from compressive measurements.


As can be observed in Fig. \ref{subfig:OAvsComp}, the proposed classification approach exhibit a competitive performance with respect the other methods. Furthermore, Fig. \ref{subfig:OAvsTraining} displays the OA curves versus the training rate for the various feature fusion approaches. For this experiment, the projections were captured by fixing the compression ratio at $25\%$. As shown in this figure, the proposed approach outperforms the other methods at low rates of training samples. 

Additionally, Fig. \ref{subfig:OAvsSNRG} shows the OA curves versus the signal-to-noise ratio (SNR) exhibited by compressive measurements for the various fusion approaches. More precisely, projections are corrupted with additive noise obeying to a zero-mean Gaussian model. For these curves, five realizations are averaged. Notice that the proposed approach exhibits a superior performance compared to the remaining methods, with at least $4\%$ of accuracy gain. In practical scenarios, detectors are affected by signal-dependent noise, therefore, we test the proposed approach when projections are corrupted by Poisson noise. In this regard, Fig. \ref{subfig:OAvsSNRP} displays the OA curves versus the Poisson noise level. As can be seen in this figure, the proposed fusion strategy outperforms the other techniques for the interval under test.

\begin{figure}
\begin{center}
\begin{tabular}{c c}
    \hspace{-55pt}
    \begin{subfigure}[b]{.40\textwidth}
    \begin{center}
    	\begin{tikzpicture}
		\begin{axis}[
		legend cell align = left,
		legend pos = south east,
		scale = 0.45,
		grid = major,
		ylabel = \footnotesize \textrm{Overall accuracy ($\%$)},
		ylabel style = {yshift=-0.2cm},
		ymin = 85,
		ymax = 100,
		xmin = 12.5,
		xmax = 50,
		xlabel = \footnotesize \textrm{Compression ratio ($\%$)},
		ticklabel style = {font=\scriptsize},
		]
		\addplot[color=green, smooth, thick, line width=2pt] table[x ={x},y = y1]{FigureData/OAvsCompressionSalinas_Response.dat};
		\addlegendentry{\tiny \textrm{FEF-TR}}
		\addplot[color=orange, smooth, thick, line width=2pt] table[x ={x},y = y2]{FigureData/OAvsCompressionSalinas_Response.dat};
		\addlegendentry{\tiny \textrm{FEF-L1TV}}
		\addplot[color = blue, smooth, thick, line width=2pt] table[x ={x},y = y4]{FigureData/OAvsCompressionSalinas_Response.dat};
		\addlegendentry{\tiny \textrm{Proposed}}
		\end{axis}
	\end{tikzpicture}
    
    \end{center}
    \vspace{-15pt}
    \caption{}
    \label{subfig:OAvsComp_SVMPLY}
    \end{subfigure}
    &
    \hspace{-75pt}
        \begin{subfigure}[b]{.40\textwidth}
    \begin{center}
    	\begin{tikzpicture}
		\begin{axis}[
		legend cell align = left,
		legend pos = south east,
		scale = 0.45,
		grid = major,
		ylabel = \footnotesize \textrm{Time (s)},
		ylabel style = {yshift=-0.2cm},
		ymin = 10,
		ymax = 1000,
		xmin = 12.5,
		xmax = 50,
		ymode=log,
		xlabel = \footnotesize \textrm{Compression ratio ($\%$)},
		ticklabel style = {font=\scriptsize},
		]
		\addplot[color=green, smooth, thick, line width=2pt] table[x ={x},y = y9]{FigureData/OAvsCompressionSalinas_Response.dat};
		\addplot[color=orange, smooth, thick, line width=2pt] table[x ={x},y = y10]{FigureData/OAvsCompressionSalinas_Response.dat};
		\addplot[color=blue, smooth, thick, line width=2pt] table[x ={x},y = y12]{FigureData/OAvsCompressionSalinas_Response.dat};
		\end{axis}
	\end{tikzpicture}
    
    \end{center}
    \vspace{-15pt}
    \caption{}
    \label{subfig:TimeVsComp}
    \end{subfigure}
\end{tabular}
\end{center}
\vspace{-25pt}
    \caption{Salinas Valley data set: (a) OA versus the compression ratio yielded by the different compressive fusion approaches using the same supervised classifier and (b) Computation time versus the compression ratio yielded by the different compressive fusion approaches.}
\end{figure}
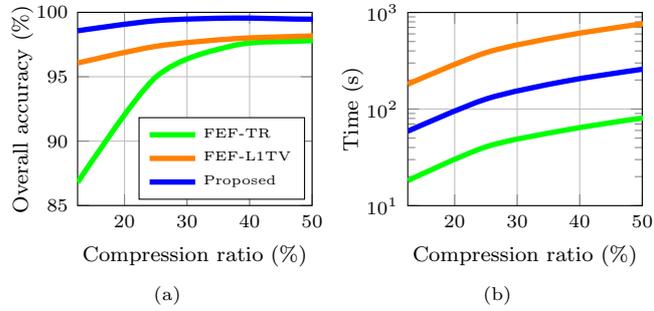

To show the performance gain of the proposed fusion method with respect to other approaches that fuse features from dual-resolution compressive measurements, Fig. \ref{subfig:OAvsComp_SVMPLY} displays the OA versus the compression ratio for the various fusion techniques using a particular supervised classifier. More precisely, we compare the proposed method with respect to FEF-TR and  FEF-L1TV methods \cite{HinojosaCoded2018,RamirezMultiresolution2019}. To this end, we use a support vector machine with a polynomial kernel as a supervised classifier. Notice that every value of this curve is obtained by averaging of ten realizations of the corresponding experiment, and for each trial, a different coded aperture pattern is generated. Furthermore, for each realization of this experiment, we selected the same 10 $\%$ of the features as training samples and the remaining 90 $\%$ as test samples. As can be seen in Fig. \ref{subfig:OAvsComp_SVMPLY}, the proposed feature fusion approach outperforms the other methods, exhibiting a remarkable accuracy gain for high compression rates. Finally, Fig. \ref{subfig:TimeVsComp} shows the computation time spent by the various feature fusion methods as the compression rate increases. These values were obtained using a desktop architecture with an Intel Core i7 CPU, 3.00 GHz, 64-GB RAM, and Ubuntu 18.04 operating system. As can be seen in this figure, the proposed fusion approach exhibits about ten times lower computation times than those spent by the FEF-L1TV method. Additionally, the proposed feature fusion approach shows the best trade-off between labeling accuracy and computation time.

\begin{figure}
\begin{center}
\begin{tabular}{cccc}
    \begin{subfigure}[b]{.20\textwidth}
    \begin{center}
        \includegraphics[width=\textwidth]{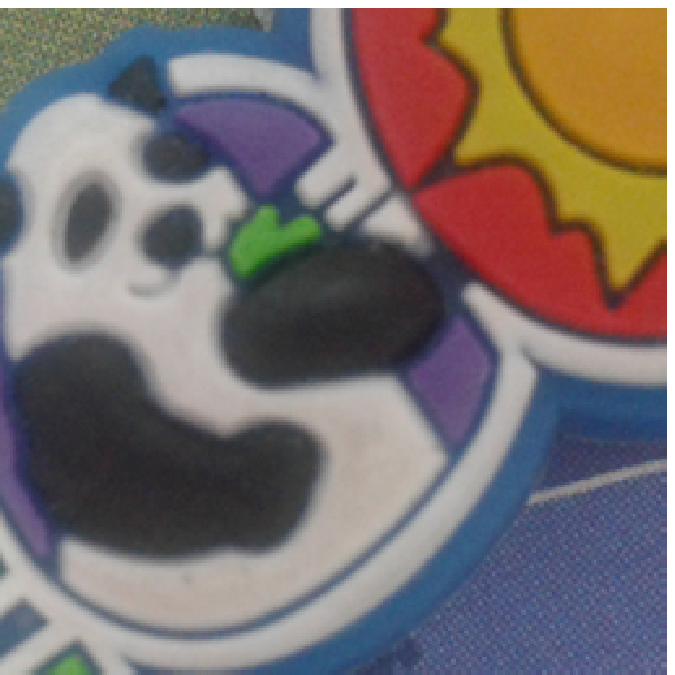}
    \end{center}
    \vspace{-15pt}
    \caption{}
    \label{subfig:rgb_realdata}
    \end{subfigure}
    &
    \begin{subfigure}[b]{.20\textwidth}
    \begin{center}
        \includegraphics[width=\textwidth]{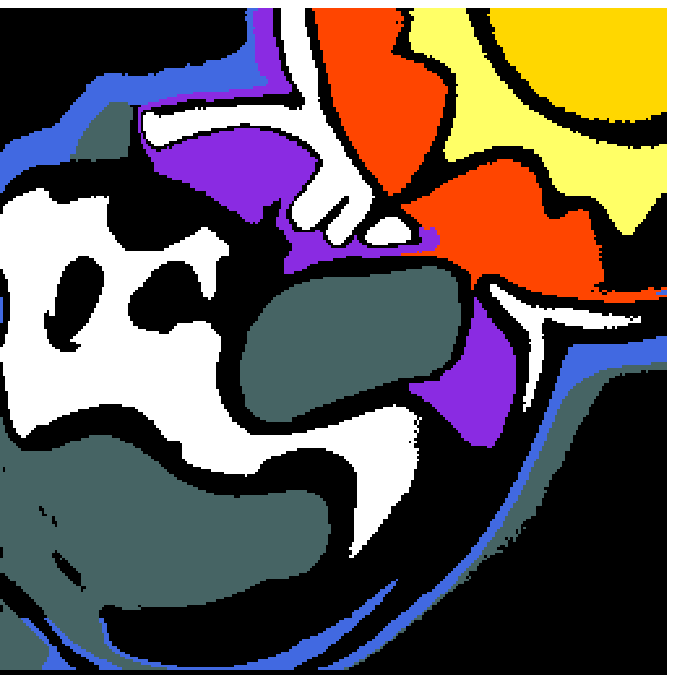}
    \end{center}
    \vspace{-15pt}
    \caption{}
    \label{subfig:gt_realdata}
    \end{subfigure}
    &
    \begin{subfigure}[b]{.20\textwidth}
    \begin{center}
        \includegraphics[width=\textwidth]{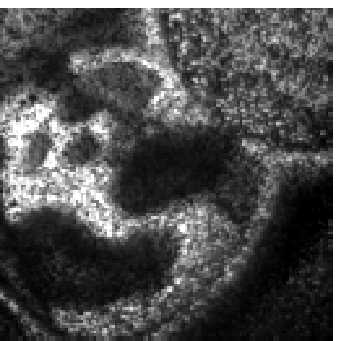}
    \end{center}
    \vspace{-15pt}
    \caption{}
    \label{subfig:csi_hs_HR}
    \end{subfigure}
    &
    \begin{subfigure}[b]{.20\textwidth}
    \begin{center}
        \includegraphics[width=\textwidth]{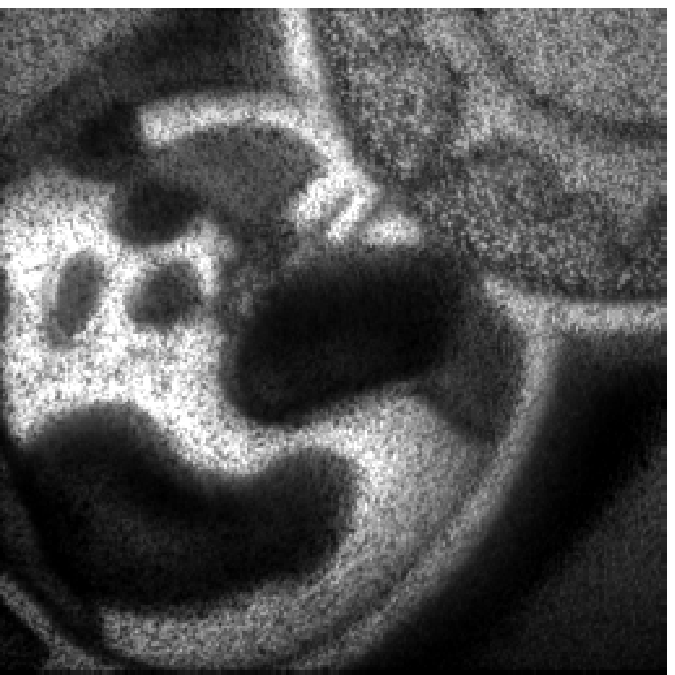}
    \end{center}
    \vspace{-15pt}
    \caption{}
    \label{subfig:csi_ms_OTVCA}
    \end{subfigure}
    \\
    \begin{subfigure}[b]{.20\textwidth}
    \begin{center}
        \includegraphics[width=\textwidth]{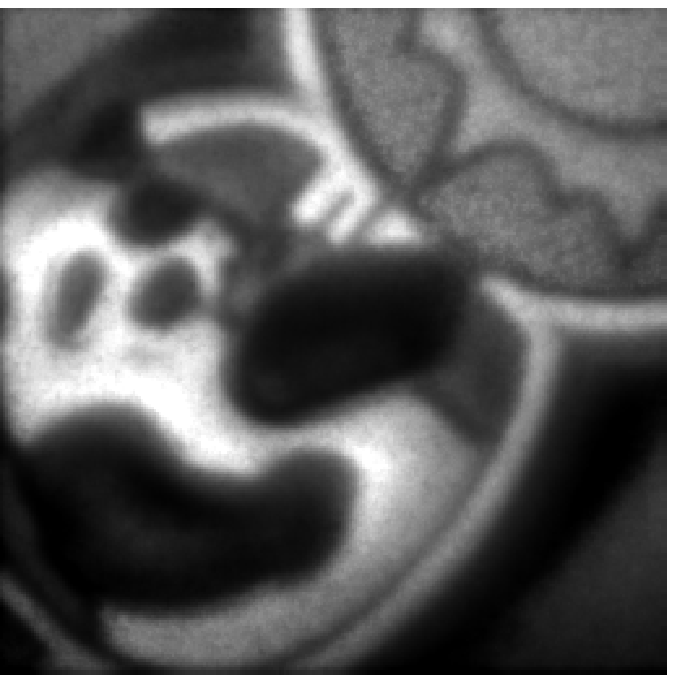}
    \end{center}
    \vspace{-15pt}
    \caption{}
    \label{subfig:band_fef_tr}
    \end{subfigure}
    &
    \begin{subfigure}[b]{.20\textwidth}
    \begin{center}
        \includegraphics[width=\textwidth]{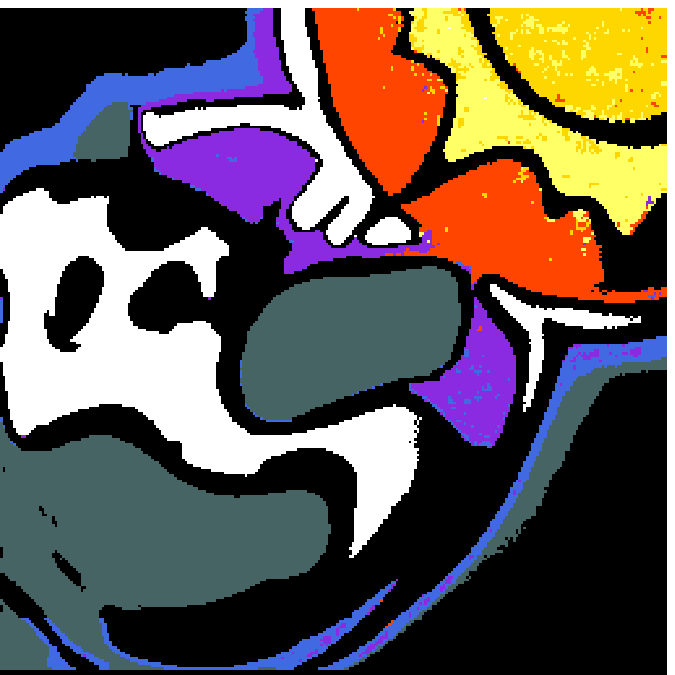}
    \end{center}
    \vspace{-15pt}
    \caption{}
    \label{subfig:cmr_fef_tr}
    \end{subfigure}
    &
    \begin{subfigure}[b]{.20\textwidth}
    \begin{center}
        \includegraphics[width=\textwidth]{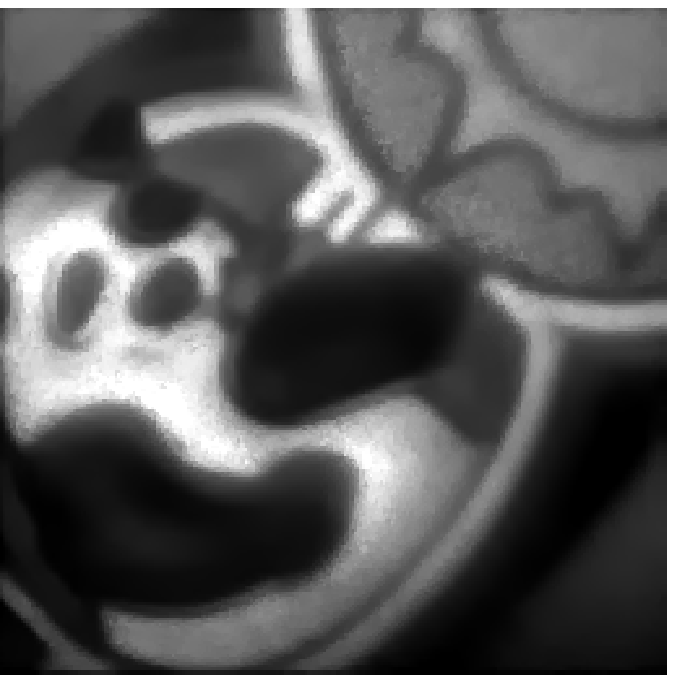}
    \end{center}
    \vspace{-15pt}
    \caption{}
    \label{subfig:band_proposed}
    \end{subfigure}
    &
    \begin{subfigure}[b]{.20\textwidth}
    \begin{center}
        \includegraphics[width=\textwidth]{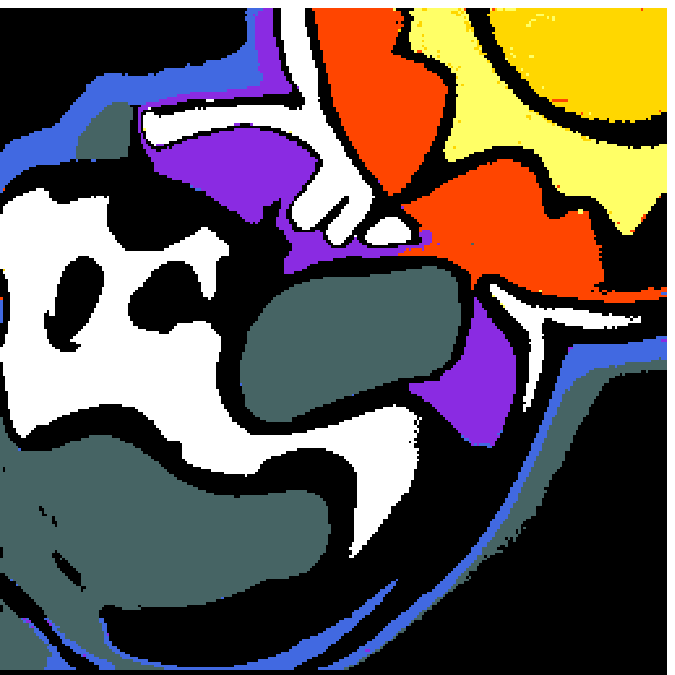}
    \end{center}
    \vspace{-15pt}
    \caption{}
    \label{subfig:cmr_proposed}
    \end{subfigure}
\end{tabular}
\end{center}
    \vspace{-10pt}
    \caption{Real CSI measurements. (a) the RGB image of the scene and (b) the classification map of the ground truth; compressive projections captured by (c) a CHSI sensor and (d) a CMSI sensor; (e) and (f) a feature band and the labeling map obtained by the FEF-TR method \cite{RamirezSpectral2020}, OA: $96.03$; (g) and (h) a feature band and the classification map obtained by the proposed approach, OA: $98.88$ $\%$.}\vspace{-10pt}
    \label{fig:my_label}
\end{figure}

\subsection{Real measurements}

Finally, the performance of the proposed classification approach from real compressive measurements is evaluated. To this end, we use data captured by a 3D-CASSI optical setup. More precisely, this data set was captured in the optics laboratory of the High-Dimensional Signal Processing (HDSP) group at the Universidad Industrial de Santander, Colombia. Furthermore, this dataset comprises 32 multispectral snapshots with dimensions $256 \times 256$ and 8 hyperspectral snapshots with dimensions $128 \times 128$. Notice that these measurements were acquired to evaluate the performance of the FEF-TR method \cite{RamirezSpectral2020}. Figure \ref{subfig:rgb_realdata} shows the RGB image of the scene and the ground truth map is displayed in Fig. \ref{subfig:gt_realdata}. Additionally, Figs \ref{subfig:csi_ms_OTVCA} and \ref{subfig:csi_hs_HR} illustrate the camera snapshots captured by both the hyperspectral CSI sensor the multispectral CSI system, respectively. 

For comparison purposes, Figs \ref{subfig:band_fef_tr} and \ref{subfig:cmr_fef_tr} illustrate a feature band and the classification map obtained by the FEF-TR approach \cite{RamirezSpectral2020}. A feature band and the labeling map estimated by the proposed classification approach are shown in Figs  \ref{subfig:band_proposed} and \ref{subfig:cmr_proposed}, respectively. Note that the proposed approach reduces the undesirable artifacts due to the acquisition noise. In addition, the proposed labeling approach remarkably minimize the classification noise providing the best overall accuracy value. 

\section{Conclusions}
\label{sec:conclu}

In this paper, a method that fuses features directly from compressive data has been developed for spectral image classification. More precisely, this approach avoids the feature extraction stage by including the information of the coded aperture patterns to obtain projection matrices. The discrete mathematical model that characterizes the compressive measurements as degraded versions of the high-resolution features was presented. Subsequently, the feature fusion was formulated as a least-squares problem regularized by two penalty terms: a sparsity-promoting term and a total variation (TV) term. Additionally, an accelerated ADMM-based algorithm has been included to numerically solve the formulated problem. The proposed feature fusion method was incorporated into an approach that classifies spectral images from compressive data. The performance of the proposed classification technique was evaluated on four spectral image data sets under different criteria. Notice that the proposed approach was successfully tested on real compressive measurements. The proposed classification approach exhibited a superior performance with respect to other state-of-the-art methods that obtain features from compressive measurements.

\section{Acknowledgments}

This project has received funding from the European Union’s Horizon 2020 research and innovation programme under the Marie Skłodowska-Curie grant agreement No 754382, GOT ENERGY TALENT. The content of this article does not reflect the official opinion of the European Union. Responsibility for the information and views expressed herein lies entirely with the authors.






\bibliographystyle{elsarticle-num-names}
\bibliography{references.bib}







\end{document}